\documentclass[12pt]{article}
\usepackage{geometry}
\usepackage[utf8]{inputenc}
\usepackage[T1]{fontenc}
\usepackage{authblk}
\usepackage[per-mode=symbol, exponent-product = \cdot]{siunitx}
\usepackage{physics} 
\usepackage{bm}
\usepackage{amsmath}
\usepackage{cases}
\usepackage{amssymb}
\usepackage{float}
\usepackage{empheq}
\usepackage{hyperref}
\usepackage{cleveref}
\usepackage{comment}
\usepackage{graphicx} 
\usepackage{caption}
\usepackage{subcaption}
\usepackage{algorithm}
\usepackage{algorithmic}
\usepackage{mathrsfs}  
\usepackage{placeins}
\usepackage{booktabs}
\usepackage{multirow}
\usepackage{pdflscape}
\usepackage{amssymb}
\usepackage{xcolor}

\DeclareSIUnit\mmhg{mmHg}

\newcommand{\uale}{\bm u^\mathrm{ALE}}
\newcommand{\df}{\bm d^\mathrm{f}}
\newcommand{\xhat}{{\widehat{\bm x}}}
\renewcommand{\k}{{\mathrm{k}}}

\newcommand{\uf}{\bm u^\mathrm{f}}
\newcommand{\pf}{p^\mathrm{f}}
\newcommand{\Omegaf}{\Omega^\mathrm{f}}
\newcommand{\Omegafhat}{\widehat{\Omega}^\mathrm{f}}
\newcommand{\Gammaf}{\Gamma^\mathrm{f}}

\newcommand{\Omegad}{\Omega^\mathrm{p}}
\newcommand{\ud}{\bm u^\mathrm{p}}
\newcommand{\pd}{p^\mathrm{p}}

\newcommand{\Omegas}{\Omega^\mathrm{s}}
\newcommand{\Omegashat}{\widehat{\Omega}^\mathrm{s}}
\newcommand{\ds}{\bm d^\mathrm{s}}

\newcommand{\calcium}{[\mathrm{Ca}^{2+}]_\mathrm{i}}

\newcommand{\lifex}{\texttt{life}\textcolor{red}{$^\texttt{x}$} }

\DeclareSIUnit\mmhg{mmHg}
\DeclareSIUnit\molar{M}

\usepackage[style=phys, backend=bibtex]{biblatex} 
\title{A comprehensive mathematical model for
cardiac perfusion}

\author[1,*]{Alberto Zingaro}
\author[2]{Christian Vergara}
\author[1]{Luca Dede'}
\author[1]{Francesco Regazzoni}
\author[1,3]{Alfio Quarteroni}
\affil[1]{\small MOX, Laboratory of Modeling and Scientific Computing, Dipartimento di Matematica, Politecnico di Milano, Piazza Leonardo da Vinci 32, 20133, Milano, Italy}
\affil[2]{LaBS, Dipartimento di Chimica, Materiali e Ingegneria Chimica “Giulio Natta”, Politecnico di Milano, Piazza Leonardo da Vinci 32, 20133, Milano, Italy}
\affil[3]{Institute of Mathematics, \'Ecole Polytechnique F\'ed\'erale de Lausanne, Station 8, Av. Piccard, CH-1015 Lausanne, Switzerland  (Professor Emeritus).}

\affil[*]{alberto.zingaro@polimi.it}

\begin{document}

\flushbottom
\maketitle
%
%

\begin{abstract}
 We present a novel mathematical model that simulates myocardial blood perfusion by  embedding   multiscale and multiphysics  features.  Our model incorporates cardiac electrophysiology, active and passive mechanics, hemodynamics, reduced valve modeling, and a multicompartment Darcy model  of perfusion . We consider a fully coupled electromechanical model of the left heart that provides input for a fully coupled Navier-Stokes – Darcy Model for myocardial perfusion. The fluid dynamics problem is modeled in a left heart geometry that includes large epicardial coronaries, while the multicompartment Darcy model is set in a biventricular domain.  Using a realistic and detailed cardiac geometry, our simulations demonstrate the accuracy of our model in describing cardiac perfusion, including myocardial blood flow maps. Additionally, we investigate the impact of a regurgitant aortic valve on myocardial perfusion, and our results indicate a reduction in myocardial perfusion due to blood flow  taken away   by the left ventricle during diastole.   To the best of our knowledge,  our work represents the first instance  where   electromechanics, hemodynamics, and perfusion are integrated into a single computational framework.
\end{abstract}

\clearpage

\section*{Introduction}

\begin{figure}[t]
	\centering
	\includegraphics[trim={1.5cm 1 1.5cm 1},clip,width = \textwidth]{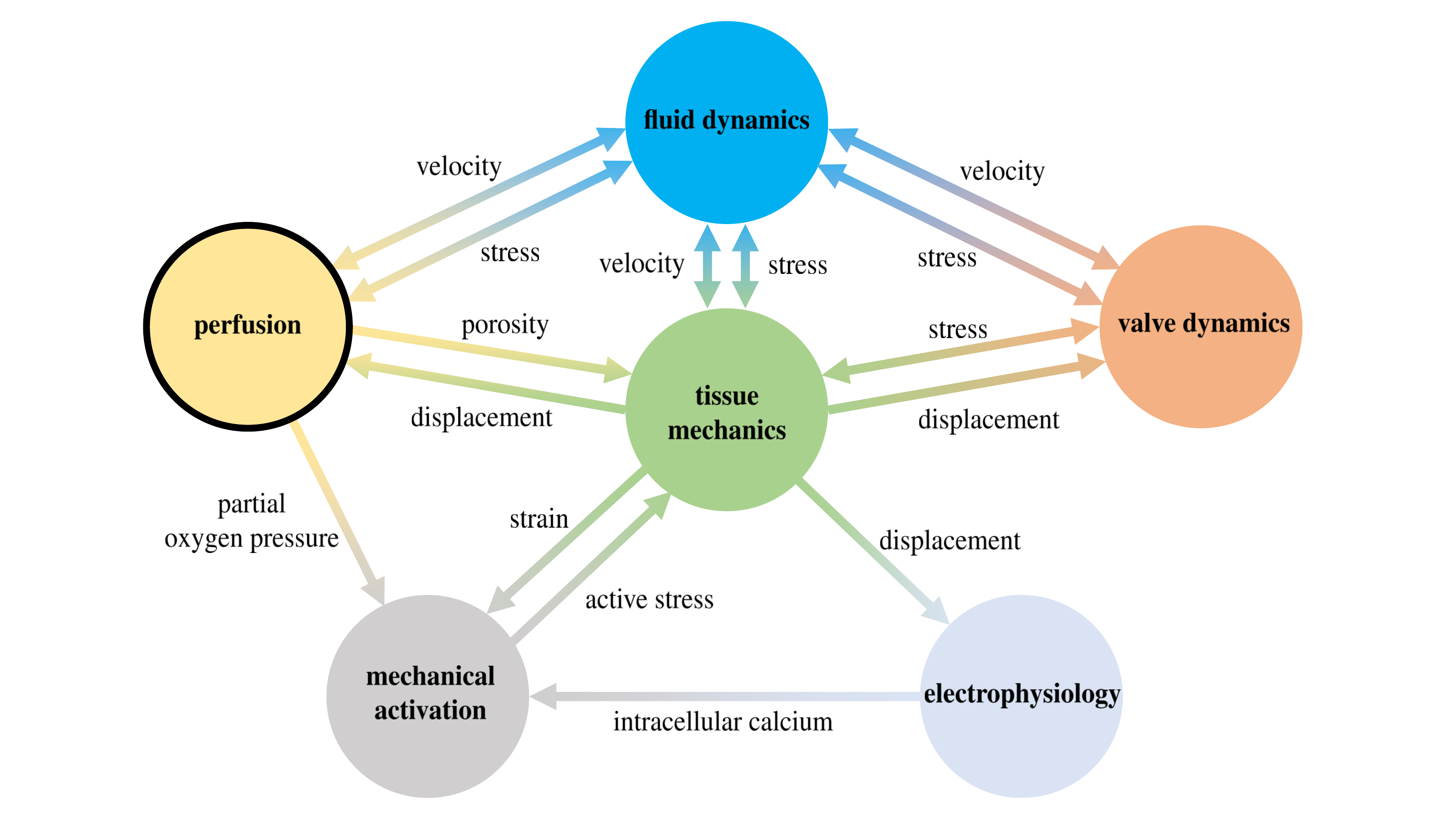}
	\caption{The perfusion is the result of complex interactions among different {models}.}
	\label{fig:interactions-in-theory}
\end{figure} 

Myocardial perfusion is the process by which oxygenated blood is delivered through the coronary arteries to the heart muscle or myocardium, enabling its oxygenation and metabolism. The microvasculature of the myocardium is responsible for facilitating the exchange of oxygen and nutrients with the blood. However, when the coronary circulation is obstructed due to factors such as arterial stenosis or cardiac pathologies like aortic regurgitation and arrhythmias, the blood supply to the cardiac muscle may be limited. This restricted blood flow can lead to ischemia and potentially trigger a myocardial infarction, commonly known as a heart attack\cite{spaan2008coronary}.

Stress myocardial computed tomography perfusion (stress-CTP) is a method for quantitatively assessing myocardial blood perfusion through myocardial blood flow maps (MBF), obtained by exposing patients to additional radiation (with respect to standard angiography) and administering an intravenous stressor during a CT scan. \textit{In-silico computational models} \cite{vankan1997finite, vankana1998mechanical, huyghe1995finite} can provide valuable insights into physiological processes and enable the simulation of virtual scenarios under multiple pathological conditions, making them useful for studying e.g. coronary by-passes \cite{guerciotti2017computational} and ventricular hypertrophy \cite{fumagalli2020image}.
However, the development of a comprehensive model of myocardial  perfusion requires accounting for the complex interactions among multiple physical processes, including the coexistence of multiple spatial scales in the coronary circulation.
The coronary arterial tree can be subdivided into \textit{epicardial coronary arteries} (large coronaries) and \textit{intramural vessels} (arterioles, venules and microvasculature) \cite{lee2012multi}. From a modeling point of view, the blood flow in the large epicardial vessels can be described using full 3D fluid dynamics or fluid-structure-interaction equations \cite{kim2010patient, lee2012multi, sankaran2012patient, kung2014vitro, sengupta2012image, schwarz2023beyond},  or  geometrically reduced hemodynamics model, as 1D models \cite{papamanolis2021myocardial, smith2002anatomically, formaggia2003one}. Differently, below a threshold length scale, the blood flow in the myocardium can be  represented as in a
 porous medium\cite{chabiniok2016multiphysics}, thanks  to Darcy or multicompartment Darcy models \cite{huyghe1995finite,michler2013computationally, di2021computational, hyde2014multi, papamanolis2021myocardial, barnafi2022multiscale}.  The integration of these models yields a coupled mathematical problem, featuring dynamic and kinematic conditions at the interface between large coronaries and microvasculature \cite{michler2013computationally, di2021computational}. 

\Cref{fig:interactions-in-theory} displays the intricate processes involved in myocardial perfusion, which result from the interplay of various physical phenomena and scales, including electrophysiology, mechanical activation, tissue mechanics, cardiac hemodynamics, and valve dynamics. In this paper, we propose for the first time a novel mathematical model that unifies these different aspects within a single framework. Our mathematical model includes \textit{core models} for electrophysiology, active and passive mechanics, blood fluid dynamics in the left atrium, ventricle, and aorta, mitral and aortic valve dynamics, and myocardial blood perfusion. To partially decouple the problem, we use a fully-coupled electromechanical model to trigger a fully coupled Navier-Stokes - multi-compartment Darcy perfusion model. To the best of our knowledge, this work represents the first attempt in the literature to integrate electromechanics, fluid mechanics, and perfusion into a single computational framework.

Our computational model provides physiological coronary flow rates and myocardial blood flow maps for the healthy case, as demonstrated by our results. In addition, we successfully simulate a severe aortic valve regurgitation, which can cause reduced oxygen delivery to the myocardium due to steal of coronary flow during diastole.

Our novel integrated model is mathematically sound and physiologically accurate, as it does not require any assumptions about boundary conditions at the inlet sections of large coronary arteries -- as commonly done for instance in refs. \cite{di2021computational, sun2014computational, zhong2018application, athani2021two, di2022prediction} --  and features a detailed 3D electro-mechano-fluid model to provide precise inputs for myocardial perfusion. Our computational model enables the simulation of the effects of various pathologies on perfusion, as demonstrated in our study on aortic regurgitation. Our work is a significant advancement towards the realization of an integrated model of the whole human heart function,  which would enable in-depth investigations of physiological and pathological perfusion scenarios, including the myocardial ischemia resulting from a coronary artery occlusion.

\section*{Methods}
To describe the methodology that we develop for the multiphysics simulation of cardiac perfusion,  we  first introduce our mathematical model, then we give details on numerical methods, software libraries, and computational setup. 

\subsection*{Mathematical model}
\begin{figure}[t]
	\centering
	\includegraphics[trim={1.5cm 1 1.5cm 1},clip,width =  \textwidth]{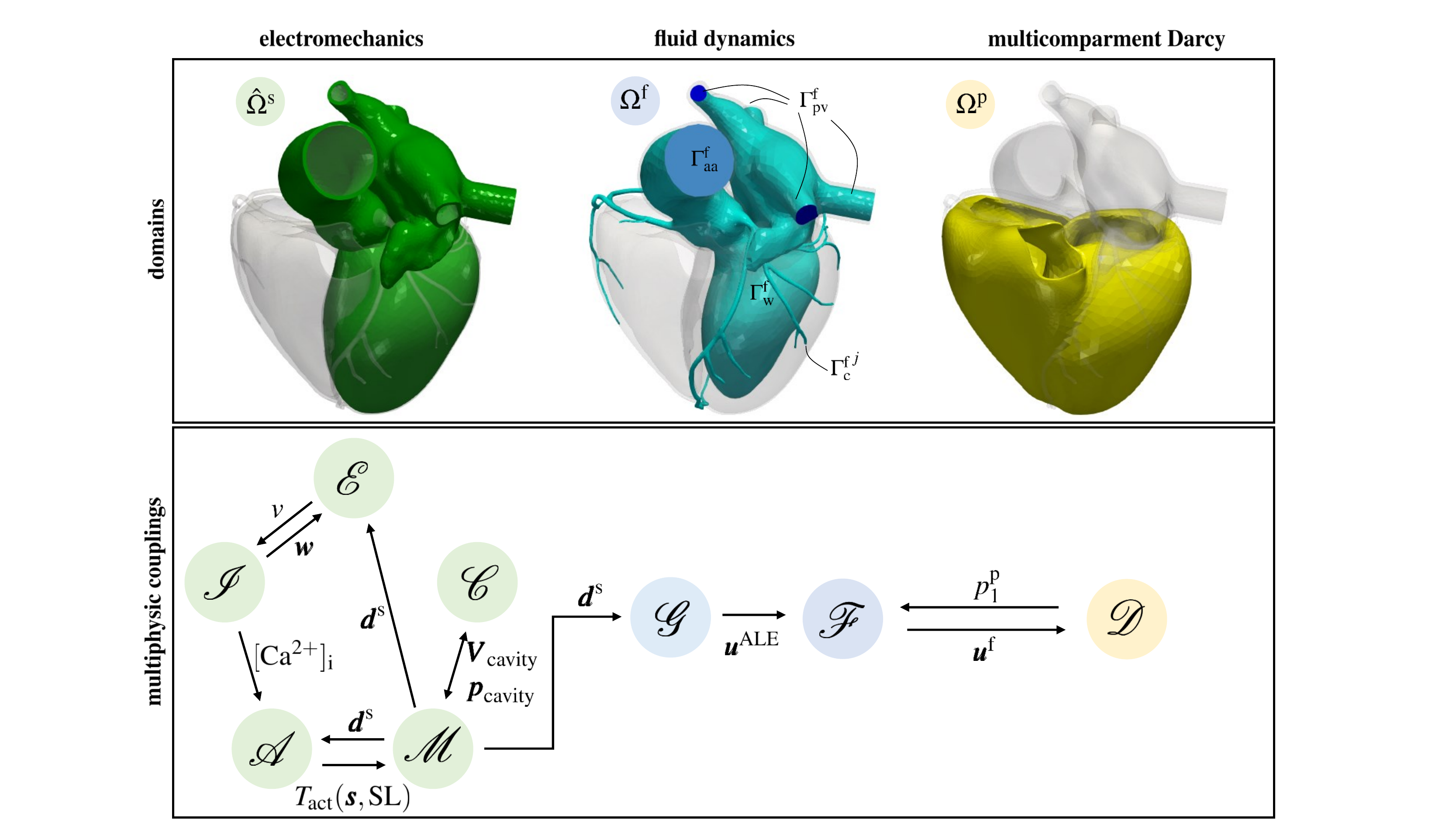}
	\caption{Top: Computational domains; Bottom: Coupling of the different physics.}
	\label{fig:coupling}
\end{figure}

For the mathematical model, we consider the time domain $(0, T)$ and three different spatial domains: 
\begin{itemize}
    \item The \textit{left heart solid domain} $\Omegas$, comprising the ventricle and atrial myocardium, and by the aortic vessel wall. {In } $\Omegas$, we define the electromechanical problem. We consider a Lagrangian framework set in the reference unloaded configuration $\Omegashat$. 
    \item The \textit{left heart fluid domain} $\Omega^{\mathrm{f}}$, comprising  the left ventricle and atrium chambers, together with the aorta and the epicardial coronaries. In  $\Omega^{\mathrm{f}}$, we define the fluid dynamics problem. $\Omega^{\mathrm{f}}$ is a time dependent domain, but we omit the subscript $t$ to keep the notation simpler. 
    \item The \textit{biventricular geometry} $\Omegad$,  {that we model as if it were } a porous medium, where we set our perfusion model. We assume $\Omegad$ to be {non deformable}\cite{michler2013computationally, hyde2014multi, di2021computational}. 
\end{itemize}
We give a graphical representation of each domain in \Cref{fig:coupling}, top. Notice that we ignore fluid dynamics in the right heart since coronaries originate from the left heart. {Hence}, there is not a direct feedback of the right heart hemodynamics on myocardial perfusion. Accordingly, also the electromechanical simulation has been performed in the left heart solely. In the following, we describe each physical problem occurring in the different domains, and provide details on the coupling conditions. The overall multiphysics model is sketched in \Cref{fig:coupling}, bottom. 

\subsubsection*{Electromechanics}
To model the electric and mechanical activity of the heart, several mathematical and numerical models have been proposed in the literature \cite{chapelle2009numerical, marx2020personalization, gurev2011models, trayanova2011electromechanical, dal2013fully, lafortune2012coupled, augustin2016anatomically, gerach2021electro}. We consider the model presented in \cite{regazzoni2022cardiac, fedele2022comprehensive} which is set in the left heart solid domain $\Omegashat$ shown in  \Cref{fig:coupling}, top. For the recovery of the reference configuration \cite{ raghavan2006non, sellier2011iterative, bols2013computational, marx2020personalization}, we refer specifically to the algorithm presented in \cite{regazzoni2022cardiac}.
We reconstruct cardiac fibers with the \textit{Laplace-Dirichlet Rule-Based Methods} \cite{bayer2005laplace, doste2019rule, bayer2012novel}, using the algorithms for ventricles and atria presented in \cite{bayer2012novel} and \cite{piersanti2021modeling}, respectively. 

We assume that the active mechanics triggered by electrophysiology is present only in the left ventricle $\Omegashat_\mathrm{LV}$. Conversely, since our focus is on the dynamics downstream the aortic valve, we treat the atrial tissue as an electrically passive material. This is a simplification that has been commonly adopted in other electromechanics \cite{augustin2016anatomically, fritz2014simulation, pfaller2019importance, strocchi2020simulating} and electro-mechano-fluid \cite{santiago2018fully,bucelli2022mathematical, sugiura2012multi} models of the heart. 

We model electrophysiology by the evolution of the transmembrane potential $v$ in the left ventricle via the \textit{monodomain equation} \cite{franzone2014mathematical}. We denote the electrophysiology problem in compact form as
\begin{equation}
	\mathcal E (v; \bm w, \bm z, \ds) = 0 \; \qquad \text{ in } \Omegashat_\mathrm{LV} \times (0, T),
	\label{eq:ep}
\end{equation}
where $\bm w$ and $\bm z$ are the gating variables and ionic concentrations, respectively. Note that the monodomain equation is augmented with \textit{mechano-electric feedbacks} \cite{salvador2022role, colli2017effects, taggart1999cardiac}, as highlighted by the dependence from the solid displacement $\ds$.
We couple \Cref{eq:ep} with the \textit{ten Tusscher and Panfilov ionic model} \cite{ten2006alternans}, that we denote in short as 
\begin{equation}
	\mathcal I ( \bm w, \bm z; v) = 0 \; \qquad \text{ in } \Omegashat_\mathrm{LV} \times (0, T).
	\label{eq:ttp}
\end{equation}
We model the active contractile force\cite{ambrosi2012active} by means of the biophysically detailed \textit{RDQ20 activation model}\cite{regazzoni2020biophysically}, which accounts for the force-sarcomere length relationship and the force-velocity relationship thanks to fiber strain-rate feedback, which we  deem essential to faithfully predict blood fluxes and velocities in the CFD simulation \cite{fedele2022comprehensive, zingaro2023electromechanics}. Denoting by $\bm s$ the state variables related to the active contractile force and by $\mathrm{SL}$ the sarcomere length, which depends on the displacement $\ds$, we express the activation model in compact notation as 
 \begin{equation}
 	\mathcal A ( \bm s; \calcium, \mathrm{SL}(\ds)) = 0 \; \qquad \text{ in } \Omegashat_\mathrm{LV} \times (0, T),
 	\label{eq:activation}
 \end{equation}
where $\calcium$ represents the intracellular calcium concentration stored in the vector function $\bm z$. Following \cite{regazzoni2020biophysically}, \Cref{eq:activation} allows then to compute the active contractile force $T_\mathrm{act}(\bm s, \mathrm{SL})$. 

For the structural problem, we consider the \textit{elastodynamic equation}, in the unknown $\ds$, in which the first Piola-Kirchhoff stress tensor is split into a passive term (depending on $\ds$ only) and an active term (depending on $\ds$ and $T_\mathrm{act}$). For the passive part, we use the \textit{Usyk anistropic strain energy function} \cite{usyk2002computational}. In short, we denote the structural problem as
 \begin{equation}
	\mathcal M (\ds;  T_\mathrm{act}, \bm p_\mathrm{cavity}) = 0 \; \qquad \text{ in } \Omegashat \times (0, T),
	\label{eq:mechanics}
\end{equation}
equipped with the following boundary conditions: generalized Robin boundary conditions \cite{regazzoni2022cardiac} to model the action of the pericardium, homogeneous Dirichlet boundary conditions (i.e. no displacement) on the rings of the pulmonary veins and homogenous Neumann boundary conditions (i.e. no stress) on the ring of the ascending aorta. Furthermore, for simplicity, we set homogenous Dirichlet boundary conditions on the epicardial valvular ring. {On the endocardium, we set the fluid pressure as described {in the following paragraph}.}

In this work, we consider a one-way coupling between the electromechanics and the 3D fluid dynamics problems \cite{augustin2016patient, karabelas2018towards, this2019augmented, zingaro2022geometric} (see below). Specifically, the 3D electromechanics problem is solved off-line, prior to the 3D fluid dynamics-perfusion problem. However, in order to account for feedback of the fluid on the electromechanical model, we strongly couple the 3D electromechanics with a 0D lumped parameter model of the circulation \cite{blanco20103d, regazzoni2022cardiac, hirschvogel2017monolithic}. Thus,  on the endocardium, we enforce the continuity of the 0D fluid - 3D solid cavity pressures and cavity volumes. Accordingly, $ \bm p_\mathrm{cavity} $ and $ \bm V_\mathrm{cavity}$  are the vectors collecting the pressures and volumes of the left atrium, left ventricle and ascending aorta. We denote the circulation model as
 \begin{equation}
	\mathcal C (\bm c, \bm V_\mathrm{cavity}^\mathrm{0D}; \bm p_\mathrm{cavity}) = 0 \; \qquad \text{ in } (0, T),
	\label{eq:circulation}
\end{equation}
where $\bm c$ is the state vector {that includes } pressures, volumes and fluxes in different compartments. Particularly, the pressure $\bm p_\mathrm{cavity}$ acts as a Lagrangian multiplier to enforce the volumetric costraint $\bm V_\mathrm{cavity}(\ds) = \bm V_\mathrm{cavity}^\mathrm{0D}$ \cite{regazzoni2022cardiac}.

\subsubsection*{The fluid geometry and fluid dynamics models}
Let $\Omegafhat \subset \mathbb{R}^3$ be the fluid dynamics domain (that is the region occupied by the fluid)  in its reference configuration (see \Cref{fig:coupling}). The fluid domain in the current configuration is 
	$\Omegaf  = \{ \bm x \in \mathbb R^3: \; \bm x = \xhat + \df(\xhat, t), \; \xhat  \in  \Omegafhat \},$
with $\df: \Omegafhat \times (0, T)$  being the domain displacement (for the sake of brevity, we omit the subscript $t$ from the fluid domain and its boundaries).  The latter is computed by solving a Laplace equation in $ {\partial \Omegafhat} \times (0, T)$ with Dirichlet boundary conditions on the physical wall: $\df = \df_{w}$ , with $\df_{w}$ equal to the electromechanical displacement $\ds$ restricted on the endocardium,  and zero on the coronaries walls (suitably smoothly merged, see \cite{fedele2021polygonal, zingaro2022geometric} for further details).
We compute the fluid domain velocity by $\uale = \pdv{\df}{t}$.
We compactly denote the \textit{fluid geometry problem} as 
\begin{equation}
	\mathscr G(\df, \uale; \bm d^{\partial \Omega}) = 0 \qquad \text{ in } \Omegafhat \times (0, T). 
	\label{eq:geometric}
\end{equation}
To model blood flows in the left heart and large epicardial coronaries, we consider the \textit{Navier-Stokes} equations expressed in \textit{Arbitrary Lagrangian Eulerian} (ALE) framework \cite{donea1982arbitrary}. We set our fluid dynamics problem in the domain $\Omegaf$, {delimited } by $\partial \Omegaf = \Gammaf_\mathrm{pv} \cup \Gammaf_\mathrm{aa} \cup \Gammaf_\mathrm{c} \cup \Gammaf_\mathrm{w}$. These boundaries represent  the pulmonary veins sections, ascending aorta section, coronary outlet sections and endocardial wall, respectively (see \Cref{fig:coupling}, top). In particular, we consider $J$ coronary outlet sections: $\Gammaf_\mathrm{c} = \cup_{j=1}^J \Gamma_\mathrm{c}^{\mathrm{f}, j}$. We denote by $\uf$ and $\pf$ the fluid velocity and pressure, respectively. We model blood as if it were a Newtonian fluid with constant density {$\rho = \SI{1.06e3}{\kilo\gram\per\metre\cubed} $} and dynamic viscosity $\mu=\SI{3.5e-3}{\kilo\gram\per\metre\per\second}$,  with the total stress tensor defined as $\sigma(\uf, \pf) = -\pf I + \mu ( \grad \uf + \grad^T \uf) $. Moreover, we account for the presence of mitral and aortic valve by means of the \textit{Resistive Immersed Implicit Surface} (RIIS) method~\cite{fedele2017patient, astorino2012robust}. We consider both valves as immersed surfaces in the fluid domain. Each valve is characterized by a resistance $R_\k$ and by a parameter $\varepsilon_\k$, representing the half-thickness of the valve leaflets. We introduce in the momentum balance a resistive term $\bm{ R}(\uf, \uale)$ that enforces the kinematic mismatch between the relative fluid velocity and the one of the valve. We refer to refs.\cite{fedele2017patient, fumagalli2020image, zingaro2022geometric, zingaro2022modeling} for further details on this method and for the definition of $\bm R$. We let the valves open and close instantaneously, at the initial and final times of isovolumetric phases (computed from the electromechanical simulation). 
The fluid dynamics model reads: 
\begin{subequations}
\label{eq:ns_ale_riis}
    \begin{empheq}[ left={ ( \mathcal F)\empheqlbrace\, } ]{alignat=3}
    & \rho \frac{\partial^\mathrm{ALE} \uf }{\partial t}  + ((\uf - \uale)\cdot \grad )\uf +  & \nonumber
    \\
    & + \div \sigma (\uf, \pf) 
	+ \bm { R}(\uf, \uale)  = \mathbf 0 & \text{in } \Omegaf \times (0, T)
     \\
    & \div \uf  = 0 &  \text{in }  \Omegaf \times (0, T),
    \\
    &-(\sigma (\uf, \pf) \bm n)\cdot \bm n + \frac{1}{\alpha_j} \int_{\Gamma_\mathrm{c}^{\mathrm{f}, j}} \uf \cdot \bm n  = p_\mathrm{c}^j & \text{on }  \Gamma_\mathrm{c}^{\mathrm{f}, j} \times(0, T), \text{ with } j = 1, \dots, J,
    \\
    & (\sigma (\uf, \pf) \bm n)\cdot \bm \tau_i = 0, \qquad i = 1, 2, & \text{ on } \Gammaf_\mathrm{c} \times(0, T),
    \\
    & \uf  =  \uale  &  \text{ on }  \Gammaf_\text{w} \times (0, T),
    \end{empheq}
\end{subequations}
where $\frac{\partial^\mathrm{ALE}}{\partial t} \bm{v} = \frac{\partial\bm{v}}{\partial t} + \left( \uale \cdot\nabla\right) \bm v$ is the ALE time derivative. 
At the wall, we prescribe the ALE velocity (computed in \Cref{eq:geometric}). On the coronary outlets, we prescribe Robin boundary conditions, where $p_{\mathrm c}^j$ is the pressure that arises from the coupling condition with the multi-comparment Darcy model and $\alpha_j$ are  conductances, with $j=1, \dots, J$ \cite{di2021computational}. Moreover, on $\Gammaf_\mathrm{c}$, we also assume null tangential tractions, being $\bm \tau_i$ two tangential vectors\cite{di2021computational}. Furthermore,	on $\Gammaf_\mathrm{pv}$ and $\Gammaf_\mathrm{aa}$, we prescribe Neumann boundary conditions. Specifically, we set a constant physiological pressure equal to 10 mmHg on the inlet pulmonary vein sections. On the outlet section of the ascending aorta, we prescribe the systemic arterial pressure resulting from the 3D-0D eletromechanical simulation. The fluid dynamics model is also equipped with a zero velocity initial condition. 

\subsubsection*{The multi-compartment Darcy model}
To model blood perfusion, we consider a multi-compartment Darcy model in the biventricular myocardial domain $\Omegad$ (see \Cref{fig:coupling}, top). This model allows us to describe several length scales featuring the myocardium and its microvasculature as a porous medium made of different compartments \cite{huyghe1995finite, hyde2013parameterisation, hyde2014multi,  di2021computational}. Specifically, we consider the three compartments Darcy equations \cite{di2021computational, hyde2014multi} in the unknown $\ud_i$, $\pd_i$, representing the Darcy velocity and pore pressure, respectively, with $i=1, 2, 3$: 
\begin{subequations}
\label{eq:multicompartment_darcy}
    \begin{empheq}[ left={ ( \mathcal D)\empheqlbrace\, } ]{alignat=4}
    & \ud_i + K_i \grad \pd_i  = \mathbf 0
	& & \text{ in }  \Omegad \times (0, T),
	\\
    &	\div \ud_i  = g_i	- \sum_{k=1}^3\beta_{i, k}  (\pd_i - \pd_k) &  \quad & \text{ in } \Omegad \times (0, T),
	\\
    &\ud_i \cdot \bm n = 0 & & \text{ on }  \partial \Omegad \times(0, T).
    \end{empheq}
\end{subequations}
$K_i$ is the permeability tensor, $g_i$ a volumetric sink term and the coefficients $\beta_{i, k}$ are the inter-compartment pressure-coupling coefficients. Following ref.\cite{di2021computational}, $g_1$ is provided by epicardial blood hemodynamics (i.e. by the coupling condition with the Navier-Stokes problem, see below) and $g_2=0$ since the second compartment does not exchange mass with the outside. Furthermore, to account for the effect of the cardiac contraction on perfusion -- still avoiding the use of a poromechanical model\cite{barnafi2022multiscale} -- 
we propose $g_3$ to surrogate the reservoir effect of the coronary bed by setting
\begin{equation}
\label{eq:bed}
g_3=-\gamma(\pd_3-p_\mathrm{bed}),\qquad p_\mathrm{bed}(t) = a_1 p_\mathrm{LV}(t) + a_2,
\end{equation}
where $\gamma$ is a suitable coefficient and the new contribution $p_\mathrm{bed}(t)$ is a function of the left ventricular pressure $p_\mathrm{LV}(t)$. The latter is obtained from the 3D-0D electromechanical problem \eqref{eq:ep}-\eqref{eq:ttp}-\eqref{eq:activation}-\eqref{eq:mechanics}-\eqref{eq:circulation}.  
In \eqref{eq:bed}, $a_1$ and $a_2$ are calibrated in order to { match physiological fluxes}.

The biventricular domain $\Omegad$ is partitioned into $J$ non-overlapping perfusion regions, such that each epicardial vessel feeds only one portion \cite{di2021computational}. For the estimation of parameters $K_i$, $\beta_{i, k}$, with $i, k = 1, 2, 3$, and for the strategy employed to partion $\Omegad$, we refer to ref.\cite{di2021computational}.

\subsubsection*{Coupling conditions}
In this section, we describe the coupling conditions enforced to match the different physics. In \Cref{fig:coupling} bottom, we sketch the overall multiphysics model and we highlight the coupling conditions. 
For the fully coupled electromechanical model, we refer the reader entirely to ref.\cite{regazzoni2022cardiac}. 

For the coupling between electromechanics and cardiac hemodynamics, we consider the following kinematic condition:
\begin{equation}
\dot{\ds} = \uf  \qquad \text{ on } \Gamma_\mathrm{w} \times (0, T),
\end{equation}
where $\ds$ is defined on the atrial and ventricular endocardium and on the endothelium of the ascending aorta. We recall that, for simplicity, we set null displacement on the coronaries wall.  

For the  fully coupled Navier-Stokes -- Darcy model, the coupling conditions read\cite{di2021computational}:
\begin{subequations}
	\begin{align}
		& p_\mathrm{c}^j = \frac{1}{|{\Omegad}^j|} \int_{{\Omegad}^j} \pd_1 \, \mathrm{d}\bm x, & &   \hspace{-3cm} \text{on } \Gamma_\mathrm{c}^{\mathrm{f}, j} \times (0, T),  \text{ with } j = 1, \dots, J, \label{eq:coupling-stress}
		\\
		 & g_1 (\bm x) = \sum_{j=1}^{J}  \frac{\chi_{{\Omegad}^j} (\bm x)}{{|{\Omegad}^j|}}
		\int_{\Gamma_\mathrm{c}^{\mathrm{f}, j}} \uf \cdot \bm n, &  & \hspace{-3cm} 
 \text{in } \Omegad \times (0, T).  \label{eq:coupling-flow}
		\end{align}
\end{subequations}
where \Cref{eq:coupling-stress} and \Cref{eq:coupling-flow} enforce the balance of internal forces and mass conservation, respectively. $\chi_{{\Omegad}^j}$ is the characteristic function of the $j$--th partition \cite{di2021computational}. 

\subsection*{Computational setup}

\begin{figure}
	\centering
	\includegraphics[trim={1 1 1 1},clip,width = \textwidth]{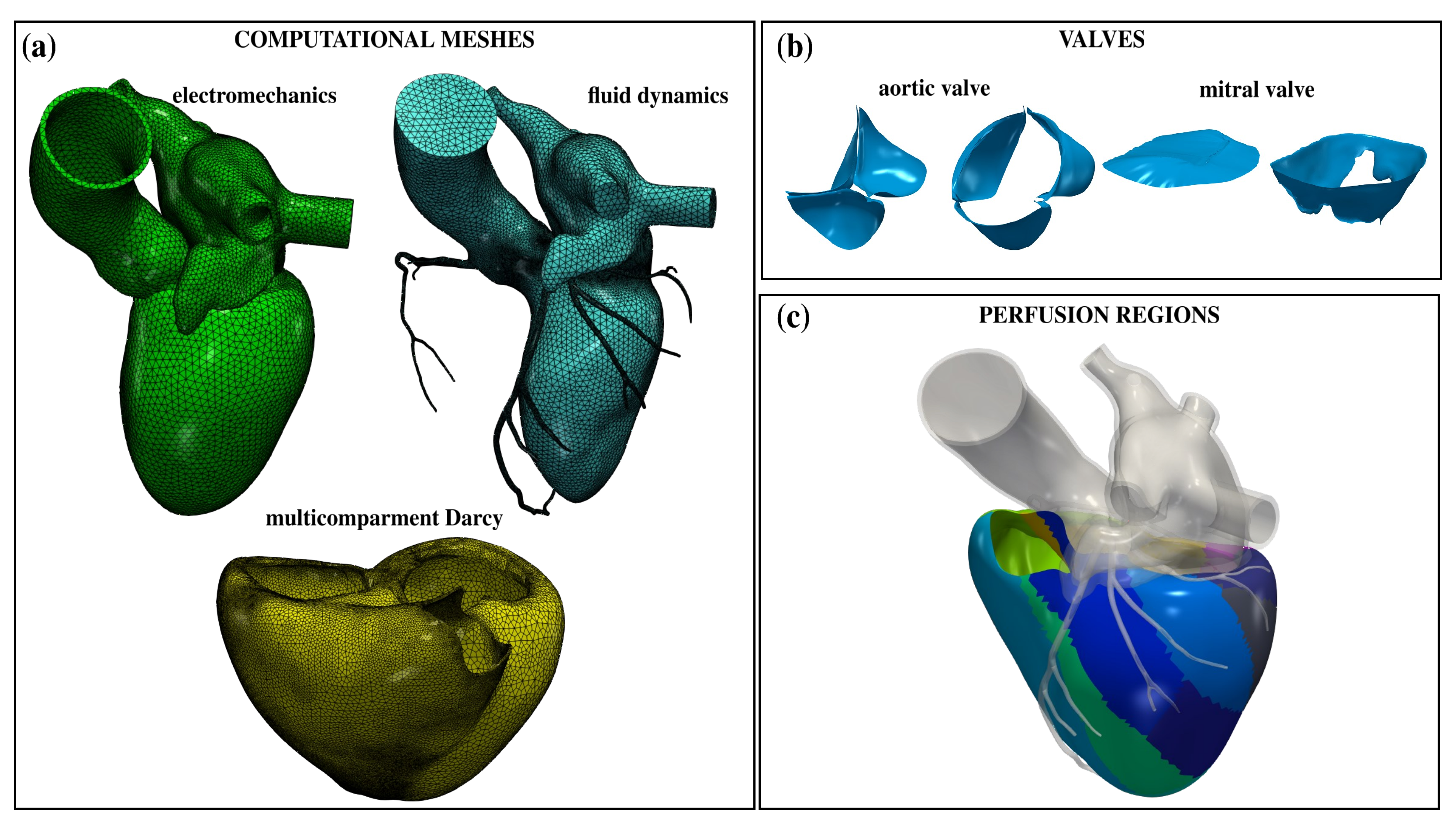}
	\caption{(a) The three computational meshes for the multiphysics problem. (b) Aortic and mitral valves in the open and closed configuration. (c) Perfusion regions in the biventricular geometry.  }
	\label{fig:mesh}
\end{figure} 

We consider a realistic cardiac geometry provided by the Zygote solid 3D heart model~\cite{zygote}: an anatomically CAD model representing an average healthy human heart reconstructed from high-resolution CT scan data. We generate three meshes for the electromechanics, fluid dynamics and multicompartment Darcy problems with \texttt{vmtk}~\cite{vmtk}, using the methods and tools discussed in~\cite{fedele2021polygonal}. Details on the generated meshes are provided in \Cref{tab:mesh} and displayed in \Cref{fig:mesh}a. Note that the CFD mesh is refined near the valves to accurately capture them with the RIIS method \cite{fedele2017patient, fumagalli2020image, zingaro2022geometric}. Immersed valves in their open and closed configurations are displayed in \Cref{fig:mesh}b. Notice also that we used the same mesh for electrophysiology and mechanics, with a value of the mesh size which is tipically considered too coarse to accurately resolve  the traveling electrical front \cite{franzone2014mathematical, quarteroni2019mathematical}. However, we suitably increase the conductivities to compensate for the use of a coarse electrophysiological mesh\cite{gerach2021electro, pezzuto2016space, woodworth2021numerical}. We believe that this choice is adequate for our purposes since the electromechanics simulation has the sole purpose to provide the endocardial displacement for the CFD problem. 

In \Cref{fig:mesh}c, we report the perfusion regions of the biventricular geometry: one for each terminal vessel. For the complete setup of the multicompartment Darcy model, 
 and for the preprocessing methods used to generate the perfusion regions, we refer the interested reader to ref.\cite{di2021computational}. To surrogate the reservoir effect of the coronary bed (see \Cref{eq:bed}), we choose $a_1 = 0.4$ and $a_2 = 1500$ Pa, which corresponds to a coronary bed pressure in the range [14.2, 61.4] mmHg.  

We discretize our mathematical models in space by the Finite Element (FE) method. We use linear FEs for electrophysiology and mechanics. The fluid dynamics problem is solved with linear FEs with VMS-LES stabilization \cite{forti2015semi, takizawa2014st}, acting also as a turbulence model to account for possible transition-to-turbulence effects \cite{zingaro2021hemodynamics}. The convective term is treated semi-implicitly. The multicompartment Darcy model, solved for the pressures, is discretized with linear FEs. As temporal advancing scheme, we use Backward Difference Formula (BDF), with the time-step sizes listed in \Cref{tab:mesh}. For additional details on numerics, we refer to refs.\cite{regazzoni2022cardiac, zingaro2022geometric, di2021computational} for the electromechanics, fluid dynamics, and perfusion models, respectively. 

 To efficiently solve the coupled problem, we first solve the electromechanical problem using a \textit{Segregated-Intergrid-Staggered} method \cite{regazzoni2022cardiac, fedele2022comprehensive}. We pick the displacement on the fifth heart cycle -- once the solution has approached a period limit cycle in terms of pressure and volume transients -- and we use it as unidirectional input\cite{zingaro2022geometric} (\textit{one-way}) for the fully coupled fluid dynamics - multicompartment Darcy problem. The electromechanical displacement is linearly projected onto the CFD wall mesh. To solve the fluid dynamics -- multicompartment Darcy problem, we use an \textit{implicit method} with an \textit{iterative splitting strategy}, i.e. we subiterate discretizations of \eqref{eq:ns_ale_riis} and \eqref{eq:multicompartment_darcy} until convergence \cite{di2021computational}. We start our simulation at the end of the filling phase. We simulate two heartbeats and we report the solution of the second  cycle to cancel the influence of a non-physical null velocity initial condition. 

We solve the multiphysics problem in \lifex \cite{africa2022flexible}, a high-performance \texttt{C++} FE library developed within the iHEART project, 
mainly focused on cardiac simulations, and based on the \texttt{deal.II} finite element core~\cite{arndt2021dealii,arndt2020dealii,dealii}. Numerical simulations are run in parallel on the GALILEO100 supercomputer (528 computing nodes each 2 x CPU Intel CascadeLake 8260, with 24 cores each, 2.4 GHz, 384GB RAM) at the CINECA supercomputing center, using 288 cores.

\begin{table}
	\centering
	\begin{tabular}{ccccccccc}
		\toprule
		\textbf{Simulation} &  {\textbf{Physics}} & \multicolumn{3}{c}{\textbf{Mesh size} {[\si{\milli\metre}]}} & {\textbf{Cells}} &  {\textbf{Vertices}}  & {\textbf{DOFs}}  & {$\Delta t$} [\si{\second}]
		\\
		& & \footnotesize{min} & \footnotesize{avg} & \footnotesize{max}  & &  & \\
		\midrule
		\multirow{3}{*}{\textbf{EM}} & $\mathcal E - \mathcal I$ & \multirow{3}{*}{\num{0.93}} & \multirow{3}{*}{\num{2.7}} & \multirow{3}{*}{\num{4.8}} & \multirow{3}{*}{\num{142056}}  &  \multirow{3}{*}{\num{31988}}   & \num{224410}& \num{1e-4}
		\\
            &  $\mathcal A$ & &  &  &  &   & \num{31988}& \num{1e-3}
            \\
		&  $\mathcal M$ & &  &  &  &   & \num{95964}& \num{1e-3}
		\\
		& $\mathcal C $ &  - & - & - & - &  -  & - &\num{1e-3}
		\\
		\midrule
            \multirow{2}{*}{\textbf{CFD-Darcy}}  & $\mathcal F$ & \num{0.03} & \num{0.92} & \num{4.03} & \num{1740644} & \num{304411} & \num{1217644} & \num{5e-4}
            \\
            &  $\mathcal D$ &  \num{0.31} & \num{1.78} & \num{5.03} & \num{214484} & \num{267374} & \num{802122} & \num{5e-4}
            \\
		\bottomrule
	\end{tabular}
	\caption{Mesh details and time step sizes for the electromechanics (EM) and computational fluid dynamics (CFD) -- Darcy simulations. Electrophysiology-ionic $\mathcal E - \mathcal I$, force generation  $\mathcal A$ and mechanics problems $\mathcal M$ are solved on the same left heart mesh. $\mathcal C$ is the 0D circulation problem. $\mathcal F$ and $\mathcal D$ are the fluid dynamics and Darcy problems, respectively.}
	\label{tab:mesh}
\end{table}

\section*{Results}

\begin{figure}
    \centering
    \includegraphics[trim={0 0 0 0},clip, width =\textwidth]{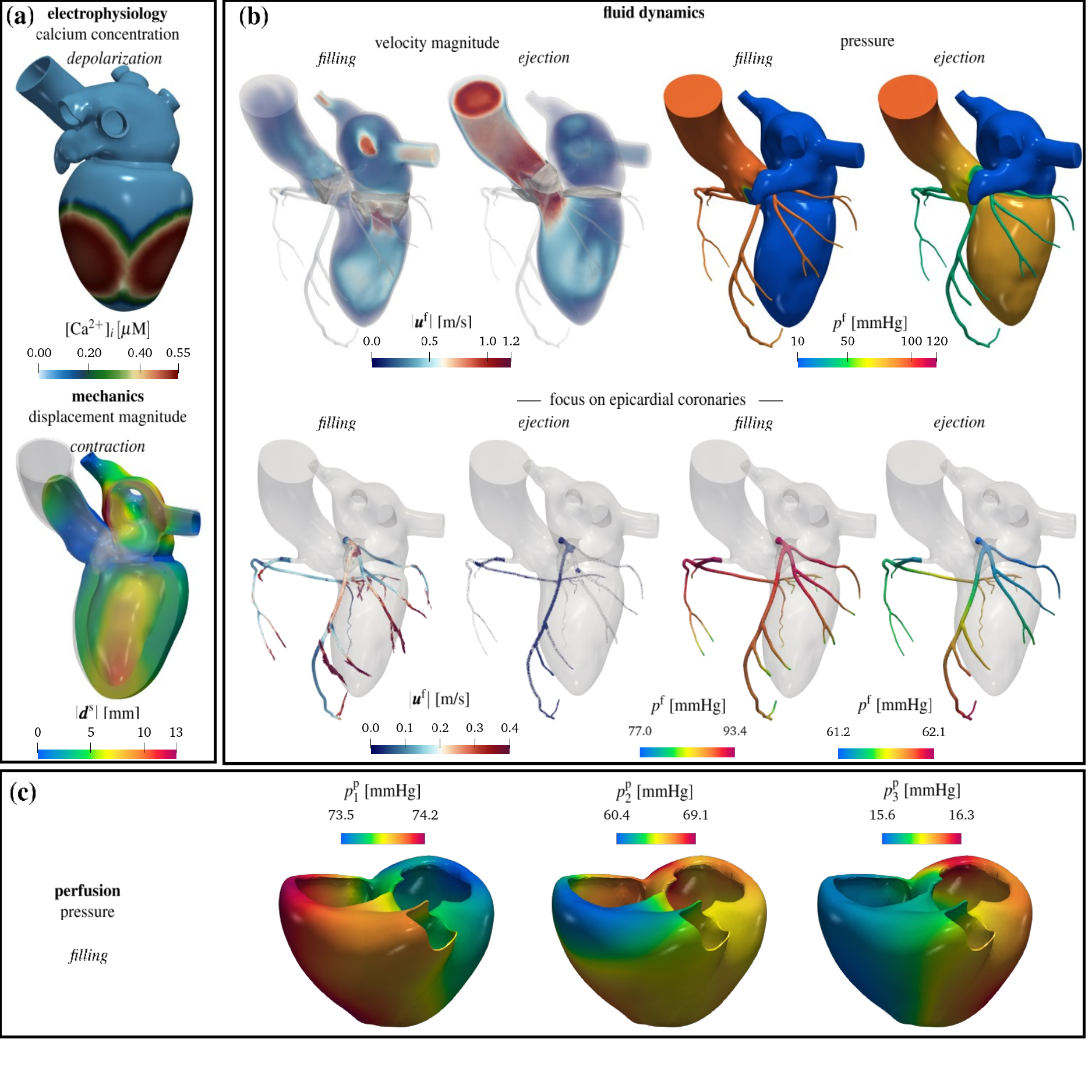}
   \caption{Results from a physiological simulation. a) electromechanics of the left heart: calcium concentration during ventricular depolarization and displacement magnitude during ventricular contraction. b) left heart hemodynamics: on the top, volume rendering of velocity magnitude and pressure during filling and ejection phases; on the bottom, focus on the epicardial coronary arteries. c) myocardial perfusion: Darcy pressure in three different compartments during filling. Test I.}
    \label{fig:results-physio-3D}
\end{figure}

We start our analysis from a physiological simulation of a coupled electromechanics - blood dynamics 
 - myocardial perfusion 
 obtained by means of the proposed multiphysics model (Test I). In \Cref{fig:results-physio-3D}, we report results from this test.
 Concerning electromechanics (\Cref{fig:results-physio-3D}(a)), we show the calcium concentration, along with the displacement magnitude when the ventricle contracts. We display the intracardiac hemodynamics  during filling and ejection in \Cref{fig:results-physio-3D}(b), by reporting the volume rendering of velocity magnitude and  pressure on the boundary of $\Omegaf$. Notice that the model can faithfully predict the formation of the clockwise jet in the left ventricle during filling, which redirects the blood in the aortic root during systole ~\cite{di2018jet, kilner2000asymmetric}. Considering the cardiac chambers only, we find larger velocities during ejection, compared to the filling phase. Conversely, focusing on the coronaries only (\Cref{fig:results-physio-3D}(b), bottom), we notice that blood is faster during the filling with respect to the  ejection phase; accordingly, a larger pressure drop is also computed  in the coronary tree during ventricular diastole, allowing the blood to accelerate and to perfuse  the cardiac muscle. In \Cref{fig:results-physio-3D}(c), we report the multicompartment Darcy's pressures during the filling stage.

\begin{figure}
    \centering
         \includegraphics[trim={0.5cm 10cm 0.5cm 10cm },clip,width=0.8\textwidth]{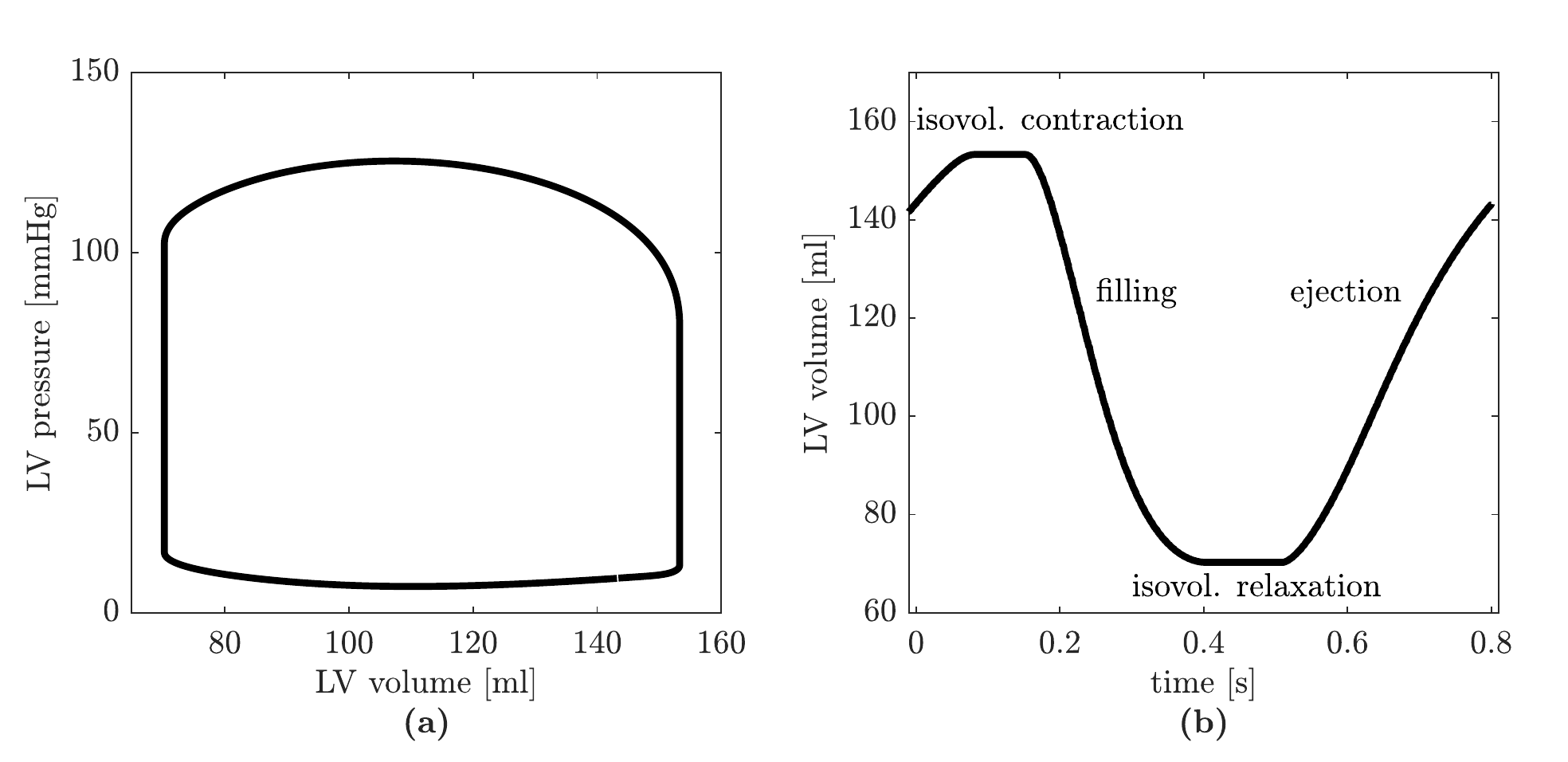}
    \caption{Results from physiological 3D electromechanical - 0D circulation simulation: (a) left ventricular pressure-volume loop; (b) left ventricular 
 volume versus time. Test I.}
    \label{fig:physiologic-em-plots}
\end{figure}

\begin{figure}
    \centering
         \includegraphics[trim={0.5cm 8.0cm 0.5cm 8.5cm },clip,width=0.8\textwidth]{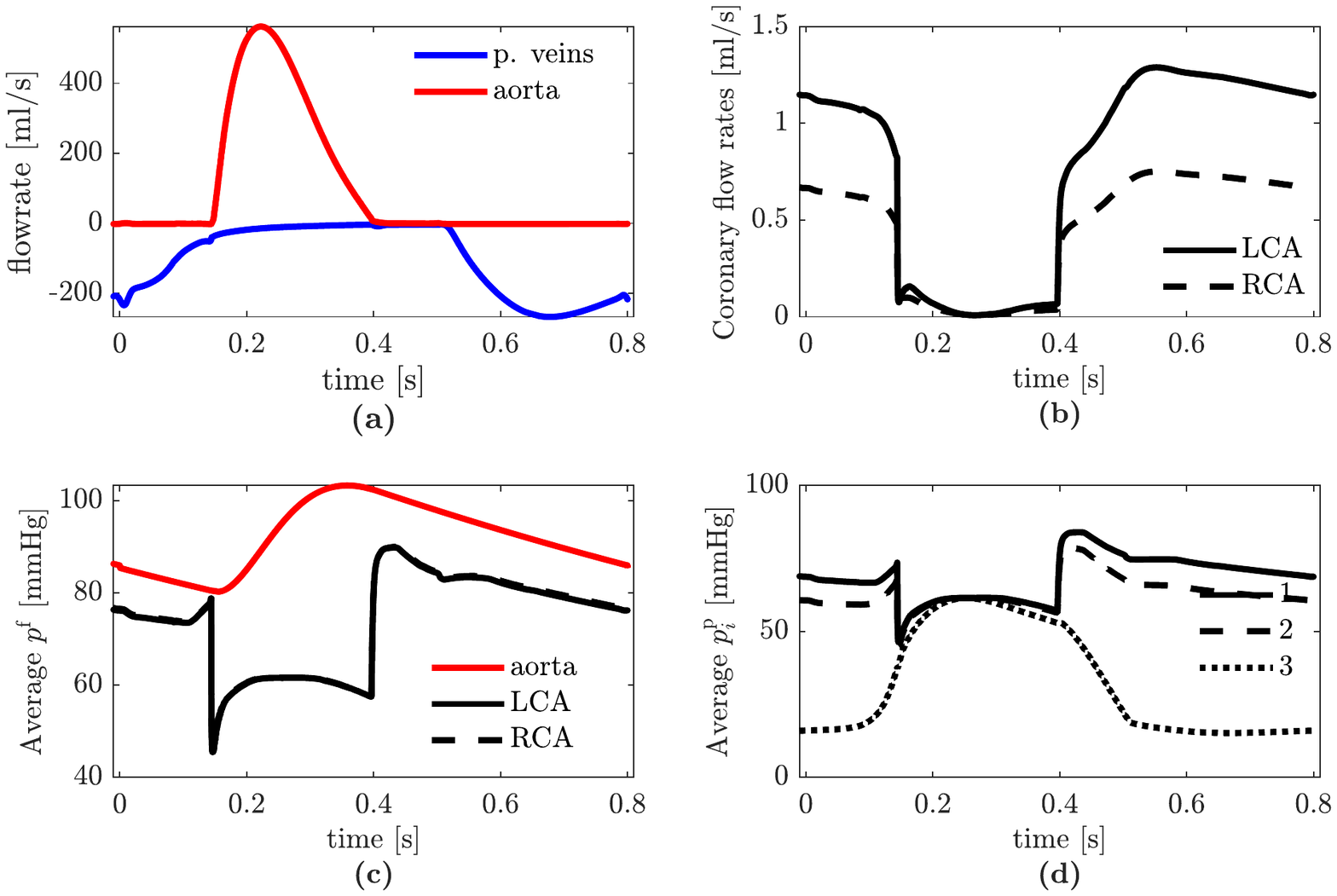}
    \caption{Results from physiological 3D Navier-Stokes -- 3D Darcy simulation in a representative heartbeat: (a) flow rates in pulmonary veins and aortic outlet section; (b) flow rates in epicardial coronary arteries; (c) average pressure in aortic outlet sections and coronary outlets; (d) pressure in the three Darcy's compartments. Test I.}
    \label{fig:physiologic-cfd-darcy-plots}
\end{figure}

\Cref{fig:physiologic-em-plots} shows quantitative results of the electromechanical simulation in Test I. Consistently with clinical ranges from literature \cite{maceira2006normalized, clay2006normal, sugimoto2017echocardiographic}, we compute the left ventricular stroke volume, ejection fraction, and peak pressure (the latter coming from the 0D hemodynamic model) equal to 83.0 ml, 54.2\%, and 125.4 mmHg, respectively (see Figures \ref{fig:physiologic-em-plots}a and \ref{fig:physiologic-em-plots}b, where pressure-volume loop and volume in time are represented). From \Cref{fig:physiologic-em-plots}b, it is possible to distinguish isovolumetric contraction, systolic ejection, isovolumetric relaxation, and diastolic filling phases. We show transients of the Navier-Stokes -- Darcy simulation in \Cref{fig:physiologic-cfd-darcy-plots}. We report  the flow rate computed at the aortic section and the total flow rate in the pulmonary veins in \Cref{fig:physiologic-cfd-darcy-plots}c: we compute a peak aortic flow equal to 562.0 ml/s -- consistently with physiological ranges \cite{hammermeister1974rate} -- and the peak total flux in the pulmonary veins is 267.2 ml/s. In \Cref{fig:physiologic-cfd-darcy-plots}d, we show the total fluxes computed at the outlets of the Left Coronary Artery (LCA) and Right Coronary Artery (RCA). Our mathematical model faithfully predicts a peak blood flow rate at the beginning of the filling phase (diastole), resulting from the myocardium relaxation after the systolic contraction. Our finding is consistent with clinical evidences \cite{johnson2008coronary}; furthermore, as also experimentally measured in  \cite{schiemann2006mr}, the flux in the LCA is larger than the one in the RCA. Pressures in the fluid dynamics domain are reported in \Cref{fig:physiologic-cfd-darcy-plots}e. We obtain a peak systolic arterial pressure of 103.3 mmHg and a minimum diastolic pressure equal to 80.3 mmHg: both results are physiologically consistent \cite{hall2020guyton}. In \Cref{fig:physiologic-cfd-darcy-plots}e, we also show the coronary pressure by averaging the average pressure in each coronary outlet: we get similar LCA and RCA pressures. \Cref{fig:physiologic-cfd-darcy-plots}f shows the pressure in the three Darcy's compartments. As expected, we have a decreasing pressure going from one compartment to the following one, and comparable values during the systolic peak, due to the contraction of the muscle and the consequent partial obstruction of vessels.

\begin{figure}
    \centering
    \includegraphics[trim={0 0 0 0},clip,width=\textwidth]{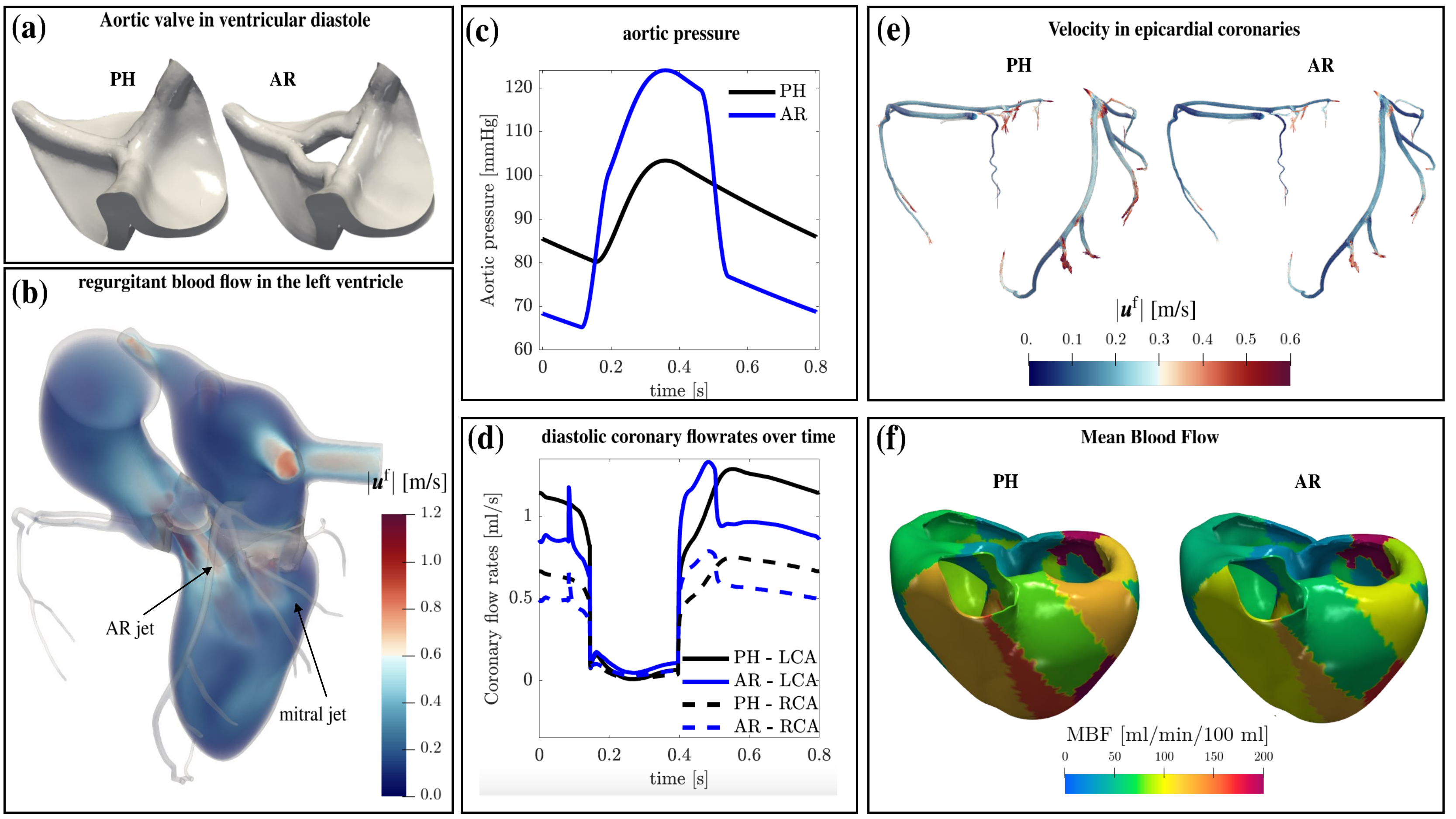}
    \caption{Simulation of Aortic Regurgitation (AR): (a) aortic valve in ventricular diastole modeled with the RIIS method, comparison between physiological (PH) and AR cases;  (b) volume rendering of velocity magnitude during ventricular filling; (c) aortic pressure prescribed on the ascending aorta outlet section in the PH and AR cases;  (d) coronary flowrates over time, comparison between PH and AR cases; (e) velocity during diastole in the coronary tree, comparison between PH and AR cases; (f) Mean Blood Flow at diastolic peak in PH and AR cases. Test II and comparisons with PH case.}
    \label{fig:results-physio-vs-patho}
\end{figure}

We aim now to study the effects that a valvular pathology has on myocardial perfusion. This allows us to explore the capabilities of the mathematical model in simulating also pathological scenarios. To this aim, we consider Test II, where the case of \textit{Aortic Regurgitation} (AR) is considered. This pathology consists of a leaking of the aortic valve leaflets causing the blood to flow from the aorta to the left ventricle during the filling stage.
To model the leaking of the aortic valve, we replace the ``closed'' physiological configuration PH used in Test I by the regurgitant configuration AR used in Test II, as we display in \Cref{fig:results-physio-vs-patho}a.
We obtain the AR configuration by introducing a regurgitant orifice which is about the 4.5\% of the aortic annulus section. Furthermore, since AR is associated with an increased systolic and a decreased diastolic aortic pressure \cite{maurer2006aortic}, we modify the systemic arterial pressure prescribed on $\Gamma_\mathrm{aa}^\mathrm{f}$ accordingly. In fact, we increase and decrease the pressure by 20\% in systole and diastole, respectively (see \Cref{fig:results-physio-vs-patho}c). \Cref{fig:results-physio-vs-patho}b shows the volume rendering of the  velocity magnitude in the AR case. During the filling stage, we observe reverse blood flow from the aorta to the left ventricle, yielding a mix of blood between the mitral and AR jets. In \Cref{fig:results-physio-vs-patho}d, we compare the coronary flowrates against time in the PH and AR cases. The diastolic flowrate decreases in the AR case, with a peak reduction of 24.8\%. This trend is also confirmed by \Cref{fig:results-physio-vs-patho}e, where we show the velocity glyphs in the coronary tree at the diastolic peak: in the AR case, we measure much lower velocities. Differently, during ejection, we observe that the AR case produces a slight increase of the coronary flow (\Cref{fig:results-physio-vs-patho}d) due to a larger systemic arterial pressure than in the PH case. To better assess the consequences of this pathology in terms of myocardial perfusion, we quantify the amount of blood inside the microvascolature. Accordingly, we compute the Myocardial Blood Flow (MBF) as: 
\begin{equation}
    \mathrm{MBF}(\bm x) = \beta_{2, 3} (\bm x) \; (\pd_2(\bm x) - \pd_3(\bm x)) 
  \; 60 [\mathrm{s}/\mathrm{min}]  \; 100 [\mathrm{ml}].
  \label{eq:mbf}
\end{equation}
MBF represents the amount of blood reaching the third compartment, i.e. where oxygen and nutrients are exchanged at a capillary level. This value is normalized over 100 ml of cardiac tissue and the factor 60 s/min allows to express the perfusion rate in minutes. \Cref{fig:results-physio-vs-patho}f shows a comparison between the PH and AR cases in terms of MBF at the diastolic peak. Overall, in both cases, we can observe a heterogeneous distribution of the MBF
due to different resistance of the vessels inside each perfusion region, provided by the heterogeneous parameters of the Darcy model \cite{di2021computational}. 
 More quantitatively, in the PH case, we compute an average MBF equal to 87.5 ml/min/100 ml. Our result is consistent with clinical studies, which measure a normal MBF at rest from 57.6 to 96.1 ml/min/100 ml \cite{kajander2011clinical}. Differently, the pathological case is characterized by a much lower perfusion: at the diastolic peak we measure 68.2 ml/min/100 ml. Thus, the ventricular reverse flow due to a regurgitant aortic valve is responsible for a steal of coronary flow, and hence an abnormal and {impaired } myocardial perfusion. 

\section*{Discussion}
In this paper, we proposed for the first time a computational model to simulate myocardial perfusion accounting for the interaction of different cardiac physical processes. Our model comprises 3D electrophysiology, active and passive mechanics, blood dynamics, and myocardial perfusion, and it was successfully applied to both a healthy and an aortic regurgitant scenarios. By carrying out simulations on a realistic cardiac geometry, we
showed that the model faithfully predicted electromechanics and blood dynamics quantities as previously shown also in \cite{regazzoni2022cardiac, zingaro2022geometric, zingaro2023electromechanics}. Moreover, as a new outcome of this work, we were able to predict cardiac perfusion in both physiological and aortic regurgitant cases by means of a comprehensive cardiac function model.

The inclusion of the whole left heart function and geometry in our simulations allowed us to avoid any  arbitrary prescription of the fluid pressure and velocities at the inlet of epicardial coronaries. Indeed, in previous perfusion models, due to the absence of any electromechanical and fluid dynamics model in the left ventricle, it was necessary to prescribe transients of flowrates or pressures at such sections  \cite{di2021computational, sun2014computational, zhong2018application, athani2021two, di2022prediction}. 
Moreover, it is well known that vascular resistance increases during systole because the contraction of the myocardium compresses the intramyocardial coronary arteries \cite{lee2012multi}, producing a peak flowrate during diastole. In our simulations, we can correctly capture this phenomenon without prescribing any data on the inlet of coronaries, but thanks to the interplay of different features that we included in the model, as discussed in what follows.
\begin{itemize}
    \item By including the entire left heart geometry and modeling its motion through the use of an electromechanics model, we were able to achieve a physiological blood velocity pattern in the ascending aorta. 
    \item  By modeling the aortic valve during the ejection phase, we were able to simulate partial obstruction of the coronary ostia when the valve is open. This approach resulted in a reduction of the systolic coronary flow rate, which is consistent with clinical evidence \cite{padula1965obstruction}. Our simulations indicated that neglecting the modeling of the aortic valve in its open configuration (i.e. by setting a null resistance during systole) leads to the computation of larger and non-physiological flow rates. For the sake of brevity, we do not include the results of this case.
    \item The contraction of cardiomyocytes is a well-known cause of impediment to systolic coronary flow \cite{katz2010physiology, vlachopoulos2011mcdonald}. To account for this effect in our perfusion model, we introduced a novel time-dependent coronary bed pressure (refer to Equation \ref{eq:bed}) that emulates an increase in vascular resistance \cite{padula1965obstruction}, thereby enabling the simulation of a diastolic coronary flow rate.
\end{itemize}

Furthermore, the development of a multiphysics mathematical  model allowed us to investigate {that a regurgitant aortic valve produces a reduction of coronary flow during diastole}, by redirecting the blood in the left ventricle, as highlighted in Figures~\ref{fig:results-physio-vs-patho}b and \ref{fig:results-physio-vs-patho}d. The main consequence of this aspect is a reduced perfusion of the myocardium during diastole, accordingly with clinical evidence \cite{rabkin2013differences} and quantified by the computation of a reduced MBF at the diastolic peak (see \Cref{fig:results-physio-vs-patho}f). Furthermore, we faithfully captured also a slight increase of the epicardial coronary flow during ejection with respect to the physiological case, as described in \cite{kume2008mechanism, rabkin2013differences} (see \Cref{fig:results-physio-vs-patho}d, blue lines during ejection). This is due to a larger aortic pressure during systole, with respect to the physiological case. In addition, the clinically meaningful outcome of the regurgitant simulation permitted us the prove the robustness of the proposed model with respect to geometric variations.

The study presented in this work features some limitations. We highlight that setting a null displacement on the epicardial valvular ring in the electromechanical simulation is not physiological. Indeed, the base of the ventricle should be free to move up and down during the cardiac cycle. Accordingly, the coronaries should follow this motion. Moreover, they should be compliant, whereas we have assumed here that they are rigid and static. 
 
{W}e noticed that the coronary pressure during ejection found by our simulation in subject PH is smaller if compared with standard physiological values and other computational studies \cite{taylor2013computational, kim2010patient}. In particular, from \Cref{fig:physiologic-cfd-darcy-plots}c, the LCA and RCA pressures should be about 25 mmHg greater during ejection. 
We believe that this is due to the open aortic valve configuration which is representative and not obtained by an FSI simulation. This limitation may result in an excessive occlusion of the coronary ostia and then in an augmented resistance during systole, which provokes a decrease of the computed coronary pressure.   

In conclusion, we expect that the incorporation of the feedback between cardiac hemodynamics and tissue mechanics through the development of a fully coupled electro-mechano-fluid-perfusion model would enable the simulation and modeling of additional pathological scenarios, such as myocardial ischemia resulting from coronary artery occlusion. We believe that the present work represents an important milestone toward the achievement of this goal. 

\printbibliography

@article{donea1982arbitrary,
	title={An arbitrary Lagrangian-Eulerian finite element method for transient dynamic fluid-structure interactions},
	author={Donea, Jean and Giuliani, S and Halleux, Jean-Pierre},
	journal={Computer Methods in Applied Mechanics and Engineering},
	volume={33},
	number={1-3},
	pages={689--723},
	year={1982},
	publisher={Elsevier}
}

@article{huyghe1995finite,
  title={Finite deformation theory of hierarchically arranged porous solids—I. Balance of mass and momentum},
  author={Huyghe, Jacques M and Van Campen, Dick H},
  journal={International Journal of Engineering Science},
  volume={33},
  number={13},
  pages={1861--1871},
  year={1995},
  publisher={Elsevier}
}

@article{guerciotti2017computational,
  title={A computational fluid--structure interaction analysis of coronary Y-grafts},
  author={Guerciotti, Bruno and Vergara, Christian and Ippolito, Sonia and Quarteroni, Alfio and Antona, Carlo and Scrofani, Roberto},
  journal={Medical {E}ngineering \& {P}hysics},
  volume={47},
  pages={117--127},
  year={2017},
  publisher={Elsevier}
}

@article{fedele2017patient,
	doi = {10.1007/s10237-017-0919-1},
	year = 2017,
	publisher = {Springer Science and Business Media {LLC}},
	volume = {16},
	number = {5},
	pages = {1779--1803},
	author = {Marco Fedele and Elena Faggiano and Luca Dede' and Alfio Quarteroni},
	title = {A patient-specific aortic valve model based on moving resistive immersed implicit surfaces},
	journal = {Biomechanics and Modeling in Mechanobiology}
}

@article{fumagalli2020image,
	doi = {10.1016/j.compbiomed.2020.103922},
	year = 2020,
	publisher = {Elsevier {BV}},
	volume = {123},
	pages = {103922},
	author = {Ivan Fumagalli and Marco Fedele and Christian Vergara and Luca Dede' and Sonia Ippolito and Francesca Nicol{\`{o}} and Carlo Antona and Roberto Scrofani and Alfio Quarteroni},
	title = {An image-based computational hemodynamics study of the Systolic Anterior Motion of the mitral valve},
	journal = {Computers in Biology and Medicine}
}

@article{zingaro2022geometric,
	title         = {A geometric multiscale model for the numerical simulation of blood flow in the human left heart}, 
	author        = {Zingaro, Alberto and Fumagalli, Ivan and Dede', Luca and Fedele, Marco and Africa, Pasquale C. and Corno, Antonio F. and Quarteroni, Alfio M.},
	year          = {2022},
	journal	= {Discrete and Continous Dynamical System - S},
	volume = {15},
	number = {8},
	pages = {2391-2427}
}

@article{zingaro2022modeling,
	title={Modeling isovolumetric phases in cardiac flows by an Augmented Resistive Immersed Implicit Surface Method},
	author={Zingaro, Alberto and Bucelli, Michele and Fumagalli, Ivan and Dede', Luca and Quarteroni, Alfio},
	journal={arXiv preprint arXiv:2208.09435},
	year={2022}
}

@article{di2021computational,
	title={A computational model applied to myocardial perfusion in the human heart: from large coronaries to microvasculature},
	author={Di Gregorio, Simone and Fedele, Marco and Pontone, Gianluca and Corno, Antonio F and Zunino, Paolo and Vergara, Christian and Quarteroni, Alfio},
	journal={Journal of Computational Physics},
	volume={424},
	pages={109836},
	year={2021},
	publisher={Elsevier}
}

@article{hyde2013parameterisation,
	title={Parameterisation of multi-scale continuum perfusion models from discrete vascular networks},
	author={Hyde, Eoin R and Michler, Christian and Lee, Jack and Cookson, Andrew N and Chabiniok, Radek and Nordsletten, David A and Smith, Nicolas P},
	journal={Medical \& Biological Engineering \& Computing},
	volume={51},
	number={5},
	pages={557--570},
	year={2013},
	publisher={Springer}
}

@article{hyde2014multi,
	title={Multi-scale parameterisation of a myocardial perfusion model using whole-organ arterial networks},
	author={Hyde, Eoin R and Cookson, Andrew N and Lee, Jack and Michler, Christian and Goyal, Ayush and Sochi, Taha and Chabiniok, Radomir and Sinclair, Matthew and Nordsletten, David A and Spaan, Jos and others},
	journal={Annals of Biomedical Engineering},
	volume={42},
	number={4},
	pages={797--811},
	year={2014},
	publisher={Springer}
}

@article{regazzoni2022cardiac,
	title={A cardiac electromechanical model coupled with a lumped-parameter model for closed-loop blood circulation},
	author={Regazzoni, F. and Salvador, M. and Africa, P.C. and Fedele, M. and Ded{\`e}, L. and Quarteroni, A.},
	journal={Journal of Computational Physics},
	volume={457},
	pages={111083},
	year={2022},
	publisher={Elsevier}
}

@article{fedele2022comprehensive,
	title={A comprehensive and biophysically detailed computational model of the whole human heart electromechanics},
	author={Fedele, Marco and Piersanti, Roberto and Regazzoni, Francesco and Salvador, Matteo and Africa, Pasquale Claudio and Bucelli, Michele and Zingaro, Alberto and Dede', Luca and Quarteroni, Alfio},
	journal={Computers Methods in Applied Mechanics and Engineering},
	volume = {410},
    pages = {115983},
    year = {2023},
    doi={https://doi.org/10.1016/j.cma.2023.115983}
}

@article{bayer2012novel,
	title={A novel rule-based algorithm for assigning myocardial fiber orientation to computational heart models},
	author={Bayer, Jason D and Blake, Robert C and Plank, Gernot and Trayanova, Natalia A},
	journal={Annals of Biomedical Engineering},
	volume={40},
	number={10},
	pages={2243--2254},
	year={2012},
	publisher={Springer}
}

@article{piersanti2021modeling,
	title={Modeling cardiac muscle fibers in ventricular and atrial electrophysiology simulations},
	author={Piersanti, Roberto and Africa, Pasquale C and Fedele, Marco and Vergara, Christian and Ded{\`e}, Luca and Corno, Antonio F and Quarteroni, Alfio},
	journal={Computer Methods in Applied Mechanics and Engineering},
	volume={373},
	pages={113468},
	year={2021},
	publisher={Elsevier}
}

@article{bucelli2022mathematical,
  title={A mathematical model that integrates cardiac electrophysiology, mechanics and fluid dynamics: application to the human left heart},
  author={Bucelli, Michele and Zingaro, Alberto and Africa, Pasquale Claudio and Fumagalli, Ivan and Dede', Luca and Quarteroni, Alfio Maria},
  journal={International Journal for Numerical Methods in Biomedical Engineering},
  pages={e3678},
  year={2022},
  publisher={Wiley Online Library}
}

@article{santiago2018fully,
	title={Fully coupled fluid-electro-mechanical model of the human heart for supercomputers},
	author={Santiago, Alfonso and Aguado-Sierra, Jazm{\'\i}n and Zavala-Ak{\'e}, Miguel and Doste-Beltran, Ruben and G{\'o}mez, Samuel and Ar{\'\i}s, Ruth and Cajas, Juan C and Casoni, Eva and V{\'a}zquez, Mariano},
	journal={International Journal for Numerical Methods in Biomedical Engineering},
	volume={34},
	number={12},
	pages={e3140},
	year={2018},
	publisher={Wiley Online Library}
}

@book{franzone2014mathematical,
	title={Mathematical cardiac electrophysiology},
	author={Franzone, Piero Colli and Pavarino, Luca Franco and Scacchi, Simone},
	volume={13},
	year={2014},
	publisher={Springer}
}

@article{salvador2022role,
	title={The role of mechano-electric feedbacks and hemodynamic coupling in scar-related ventricular tachycardia},
	author={Salvador, Matteo and Regazzoni, Francesco and Pagani, Stefano and Trayanova, Natalia and Quarteroni, Alfio and others},
	journal={Computers in Biology and Medicine},
	volume={142},
	pages={105203},
	year={2022},
	publisher={Elsevier}
}

@article{ten2006alternans,
	title={Alternans and spiral breakup in a human ventricular tissue model},
	author={Ten Tusscher, Kirsten HWJ and Panfilov, Alexander V},
	journal={American Journal of Physiology-Heart and Circulatory Physiology},
	volume={291},
	number={3},
	pages={H1088--H1100},
	year={2006},
	publisher={American Physiological Society}
}

@article{regazzoni2020biophysically,
	author =        {Regazzoni, F. and Ded{\`e}, L. and Quarteroni, A.},
	journal =       {PLOS Computational Biology},
	number =        {10},
	pages =         {e1008294},
	publisher =     {PLOS},
	title =         {Biophysically detailed mathematical models of
	multiscale cardiac active mechanics},
	volume =        {16},
	year =          {2020},
	doi =           {10.1371/journal.pcbi.1008294}
}

@article{usyk2002computational,
	title={Computational model of three-dimensional cardiac electromechanics},
	author={Usyk, Taras P and LeGrice, Ian J and McCulloch, Andrew D},
	journal={Computing and Visualization in Science},
	volume={4},
	number={4},
	pages={249--257},
	year={2002},
	publisher={Springer}
}

@article{blanco20103d,
	title={A 3D-1D-0D computational model for the entire cardiovascular system},
	author={Blanco, Pablo J and Feij{\'o}o, Ra{\'u}l A},
	journal={Mec{\'a}nica Computacional},
	volume={29},
	number={59},
	pages={5887--5911},
	year={2010}
}

@article{hirschvogel2017monolithic,
	doi = {10.1002/cnm.2842},
	year = 2017,
	publisher = {Wiley},
	volume = {33},
	number = {8},
	pages = {e2842},
	author = {Marc Hirschvogel and Marina Bassilious and Lasse Jagschies and Stephen M. Wildhirt and Michael W. Gee},
	title = {A monolithic 3D-0D coupled closed-loop model of the heart and the vascular system: Experiment-based parameter estimation for patient-specific cardiac mechanics},
	journal = {International Journal for Numerical Methods in Biomedical Engineering}
}

@article{spaan2008coronary,
  title={Coronary structure and perfusion in health and disease},
  author={Spaan, Jos and Kolyva, Christina and van den Wijngaard, Jeroen and Ter Wee, Rene and van Horssen, Pepijn and Piek, Jan and Siebes, Maria},
  journal={Philosophical Transactions of the Royal Society A: Mathematical, Physical and Engineering Sciences},
  volume={366},
  number={1878},
  pages={3137--3153},
  year={2008},
  publisher={The Royal Society London}
}

@article{lee2012multi,
  title={The multi-scale modelling of coronary blood flow},
  author={Lee, Jack and Smith, Nicolas P},
  journal={Annals of Biomedical Engineering},
  volume={40},
  number={11},
  pages={2399--2413},
  year={2012},
  publisher={Springer}
}

@article{sankaran2012patient,
  title={Patient-specific multiscale modeling of blood flow for coronary artery bypass graft surgery},
  author={Sankaran, Sethuraman and Esmaily Moghadam, Mahdi and Kahn, Andrew M and Tseng, Elaine E and Guccione, Julius M and Marsden, Alison L},
  journal={Annals of Biomedical Engineering},
  volume={40},
  number={10},
  pages={2228--2242},
  year={2012},
  publisher={Springer}
}

@article{chabiniok2016multiphysics,
  title={Multiphysics and multiscale modelling, data--model fusion and integration of organ physiology in the clinic: ventricular cardiac mechanics},
  author={Chabiniok, Radomir and Wang, Vicky Y and Hadjicharalambous, Myrianthi and Asner, Liya and Lee, Jack and Sermesant, Maxime and Kuhl, Ellen and Young, Alistair A and Moireau, Philippe and Nash, Martyn P and others},
  journal={Interface Focus},
  volume={6},
  number={2},
  pages={20150083},
  year={2016},
  publisher={The Royal Society}
}

@article{michler2013computationally,
  title={A computationally efficient framework for the simulation of cardiac perfusion using a multi-compartment Darcy porous-media flow model},
  author={Michler, Ch and Cookson, AN and Chabiniok, Radomir and Hyde, E and Lee, J and Sinclair, M and Sochi, T and Goyal, A and Vigueras, G and Nordsletten, DA and others},
  journal={International Journal for Numerical Methods in Biomedical Engineering},
  volume={29},
  number={2},
  pages={217--232},
  year={2013},
  publisher={Wiley Online Library}
}

@article{africa2022flexible,
  title={lifex: A flexible, high performance library for the numerical solution of complex finite element problems},
  author={Africa, Pasquale Claudio},
  journal={SoftwareX},
  volume={20},
  pages={101252},
  year={2022},
  publisher={Elsevier}
}

@article{ arndt2020dealii,
  title   = {The \texttt{deal.II} finite element library: design, features, and insights},
  author  = {Arndt, D. and Bangerth, W. and Davydov, D. and
  Heister, T. and Heltai, L. and Kronbichler, M. and Maier, M. and Pelteret, J. and Turcksin, B. and Wells, D.},
  journal = {Computers \& Mathematics with Applications},
  year  = {2020},
  DOI   = {10.1016/j.camwa.2020.02.022}
}

@article{arndt2021dealii,
    author      = {Arndt, Daniel and Bangerth, Wolfgang and Blais, Bruno and Fehling, Marc and Gassmöller, Rene and Heister, Timo and Heltai, Luca and Köcher, Uwe and Kronbichler, Martin and Maier, Matthias and Munch, Peter and Pelteret, Jean-Paul and Proell, Sebastian and Simon, Konrad and Turcksin, Bruno and Wells, David and Zhang, Jiaqi},
    year        = {2021},
    title       = {The deal.II library, Version 9.3},
    volume      = {29},
    journal     = {Journal of Numerical Mathematics},
}

@misc { dealii,
  title = {Official \texttt{deal.ii} website},
  howpublished = {\url{https://www.dealii.org/}},
  key = {dealii}
}

@misc{zygote,
  author  = {Zygote Media Group Inc.}, 
  title   = {Zygote solid 3D heart generation II developement report. tech. rep.},
  year    = 2014
}

@article{fedele2021polygonal,
	author 		= {Fedele, Marco and Quarteroni, Alfio M.},
	year 		= {2021},
	pages 		= {e3435},
	title 			= {Polygonal surface processing and mesh generation tools for numerical simulations of the complete cardiac function},
	journal 	= {International Journal for Numerical Methods in Biomedical Engineering},
	volume		= {37}
}

@article{vmtk,
	doi = {10.1007/s11517-008-0420-1},
	year = 2008,
	publisher = {Springer Science and Business Media {LLC}},
	volume = {46},
	number = {11},
	pages = {1097--1112},
	author = {Luca Antiga and Marina Piccinelli and Lorenzo Botti and Bogdan Ene-Iordache and Andrea Remuzzi and David A. Steinman},
	title = {An image-based modeling framework for patient-specific computational hemodynamics},
	journal = {Medical {\&} Biological Engineering {\&} Computing}
}

@article{forti2015semi,
  title={Semi-implicit {BDF} time discretization of the {N}avier--{S}tokes equations with {VMS}-{LES} modeling in a high performance computing framework},
  author={Forti, Davide and Ded{\`e}, Luca},
  journal={Computers \& Fluids},
  volume={117},
  pages={168--182},
  year={2015},
  publisher={Elsevier}
}

@article{zingaro2021hemodynamics,
  title={Hemodynamics of the heart’s left atrium based on a Variational Multiscale-LES numerical method},
  author={Zingaro, Alberto and Dede', Luca and Menghini, Filippo and Quarteroni, Alfio},
  journal={European Journal of Mechanics-B/Fluids},
  volume={89},
  pages={380--400},
  year={2021},
  publisher={Elsevier}
}

@article{zingaro2023electromechanics,
	title={An electromechanics-driven fluid dynamics model for the simulation of the whole human heart},
	author={Zingaro, Alberto and Bucelli, Michele and Piersanti, Roberto and Regazzoni, Francesco and Dede', Luca and Quarteroni, Alfio},
	journal={arXiv preprint ArXiv:2301.02148},
	year={2023}
}

@inproceedings{chapelle2009numerical,
  title={Numerical simulation of the electromechanical activity of the heart},
  author={Chapelle, Dominique and Fern{\'a}ndez, Miguel A and Gerbeau, Jean-Fr{\'e}d{\'e}ric and Moireau, Philippe and Sainte-Marie, Jacques and Zemzemi, Nejib},
  booktitle={International Conference on Functional Imaging and Modeling of the Heart},
  pages={357--365},
  year={2009},
  organization={Springer}
}

@article{marx2020personalization,
  title={Personalization of electro-mechanical models of the pressure-overloaded left ventricle: fitting of windkessel-type afterload models},
  author={Marx, Laura and Gsell, Matthias AF and Rund, Armin and Caforio, Federica and Prassl, Anton J and Toth-Gayor, Gabor and Kuehne, Titus and Augustin, Christoph M and Plank, Gernot},
  journal={Philosophical Transactions of the Royal Society A},
  volume={378},
  number={2173},
  pages={20190342},
  year={2020},
  publisher={The Royal Society Publishing}
}

@article{gurev2011models,
  title={Models of cardiac electromechanics based on individual hearts imaging data},
  author={Gurev, Viatcheslav and Lee, Ted and Constantino, Jason and Arevalo, Hermenegild and Trayanova, Natalia A},
  journal={Biomechanics and Modeling in Mechanobiology},
  volume={10},
  number={3},
  pages={295--306},
  year={2011},
  publisher={Springer}
}

@article{trayanova2011electromechanical,
  title={Electromechanical models of the ventricles},
  author={Trayanova, Natalia A and Constantino, Jason and Gurev, Viatcheslav},
  journal={American Journal of Physiology-Heart and Circulatory Physiology},
  volume={301},
  number={2},
  pages={H279--H286},
  year={2011},
  publisher={American Physiological Society Bethesda, MD}
}

@article{dal2013fully,
  title={A fully implicit finite element method for bidomain models of cardiac electromechanics},
  author={Dal, H{\"u}sn{\"u} and G{\"o}ktepe, Serdar and Kaliske, Michael and Kuhl, Ellen},
  journal={Computer Methods in Applied Mechanics and Engineering},
  volume={253},
  pages={323--336},
  year={2013},
  publisher={Elsevier}
}

@article{lafortune2012coupled,
  title={Coupled electromechanical model of the heart: parallel finite element formulation},
  author={Lafortune, Pierre and Ar{\'\i}s, Ruth and V{\'a}zquez, Mariano and Houzeaux, Guillaume},
  journal={International Journal for Numerical Methods in Biomedical Engineering},
  volume={28},
  number={1},
  pages={72--86},
  year={2012},
  publisher={Wiley Online Library}
}

@article{gerach2021electro,
  title={Electro-mechanical whole-heart digital twins: a fully coupled multi-physics approach},
  author={Gerach, Tobias and Schuler, Steffen and Fr{\"o}hlich, Jonathan and Lindner, Laura and Kovacheva, Ekaterina and Moss, Robin and W{\"u}lfers, Eike Moritz and Seemann, Gunnar and Wieners, Christian and Loewe, Axel},
  journal={Mathematics},
  volume={9},
  number={11},
  pages={1247},
  year={2021},
  publisher={MDPI}
}

@article{sellier2011iterative,
  title={An iterative method for the inverse elasto-static problem},
  author={Sellier, M},
  journal={Journal of Fluids and Structures},
  volume={27},
  number={8},
  pages={1461--1470},
  year={2011},
  publisher={Elsevier}
}

@article{bols2013computational,
  title={A computational method to assess the in vivo stresses and unloaded configuration of patient-specific blood vessels},
  author={Bols, Joris and Degroote, Joris and Trachet, Bram and Verhegghe, Benedict and Segers, Patrick and Vierendeels, Jan},
  journal={Journal of Computational and Applied mathematics},
  volume={246},
  pages={10--17},
  year={2013},
  publisher={Elsevier}
}

@article{raghavan2006non,
  title={Non-invasive determination of zero-pressure geometry of arterial aneurysms},
  author={Raghavan, ML and Ma, Baoshun and Fillinger, Mark and others},
  journal={Annals of Biomedical Engineering},
  volume={34},
  number={9},
  pages={1414--1419},
  year={2006},
  publisher={Springer}
}

@article{doste2019rule,
  title={A rule-based method to model myocardial fiber orientation in cardiac biventricular geometries with outflow tracts},
  author={Doste, Ruben and Soto-Iglesias, David and Bernardino, Gabriel and Alcaine, Alejandro and Sebastian, Rafael and Giffard-Roisin, Sophie and Sermesant, Maxime and Berruezo, Antonio and Sanchez-Quintana, Damian and Camara, Oscar},
  journal={International Journal for Numerical Methods in Biomedical Engineering},
  volume={35},
  number={4},
  pages={e3185},
  year={2019},
  publisher={Wiley Online Library}
}

@article{bayer2005laplace,
  title={Laplace--Dirichlet energy field specification for deformable models. An FEM approach to active contour fitting},
  author={Bayer, Jason D and Beaumont, Jacques and Krol, Andrzej},
  journal={Annals of Biomedical Engineering},
  volume={33},
  number={9},
  pages={1175--1186},
  year={2005},
  publisher={Springer}
}

@article{sugiura2012multi,
  title={Multi-scale simulations of cardiac electrophysiology and mechanics using the University of Tokyo heart simulator},
  author={Sugiura, Seiryo and Washio, Takumi and Hatano, Asuka and Okada, Junichi and Watanabe, Hiroshi and Hisada, Toshiaki},
  journal={Progress in Biophysics and Molecular Biology},
  volume={110},
  number={2-3},
  pages={380--389},
  year={2012},
  publisher={Elsevier}
}

@article{fritz2014simulation,
  title={Simulation of the contraction of the ventricles in a human heart model including atria and pericardium},
  author={Fritz, Thomas and Wieners, Christian and Seemann, Gunnar and Steen, Henning and D{\"o}ssel, Olaf},
  journal={Biomechanics and Modeling in Mechanobiology},
  volume={13},
  number={3},
  pages={627--641},
  year={2014},
  publisher={Springer}
}

@article{pfaller2019importance,
  title={The importance of the pericardium for cardiac biomechanics: from physiology to computational modeling},
  author={Pfaller, Martin R and H{\"o}rmann, Julia M and Weigl, Martina and Nagler, Andreas and Chabiniok, Radomir and Bertoglio, Crist{\'o}bal and Wall, Wolfgang A},
  journal={Biomechanics and Modeling in Mechanobiology},
  volume={18},
  number={2},
  pages={503--529},
  year={2019},
  publisher={Springer}
  }

@article{strocchi2020simulating,
  title={Simulating ventricular systolic motion in a four-chamber heart model with spatially varying robin boundary conditions to model the effect of the pericardium},
  author={Strocchi, Marina and Gsell, Matthias AF and Augustin, Christoph M and Razeghi, Orod and Roney, Caroline H and Prassl, Anton J and Vigmond, Edward J and Behar, Jonathan M and Gould, Justin S and Rinaldi, Christopher A and others},
  journal={Journal of Biomechanics},
  volume={101},
  pages={109645},
  year={2020},
  publisher={Elsevier}
}

@article{colli2017effects,
  title={Effects of mechanical feedback on the stability of cardiac scroll waves: A bidomain electro-mechanical simulation study},
  author={Colli Franzone, P and Pavarino, LF and Scacchi, S},
  journal={Chaos: An Interdisciplinary Journal of Nonlinear Science},
  volume={27},
  number={9},
  pages={093905},
  year={2017},
  publisher={AIP Publishing LLC}
}

@article{taggart1999cardiac,
  title={Cardiac mechano-electric feedback in man: clinical relevance},
  author={Taggart, Peter and Sutton, Peter MI},
  journal={Progress in Biophysics and Molecular Biology},
  volume={71},
  number={1},
  pages={139--154},
  year={1999},
  publisher={Elsevier}
}

@article{astorino2012robust,
  title={A robust and efficient valve model based on resistive immersed surfaces},
  author={Astorino, Matteo and Hamers, Jeroen and Shadden, Shawn C and Gerbeau, Jean-Fr{\'e}d{\'e}ric},
  journal={International Journal for Numerical Methods in Biomedical Engineering},
  volume={28},
  number={9},
  pages={937--959},
  year={2012},
  publisher={Wiley Online Library}
}

@article{takizawa2014st,
  title={ST and ALE-VMS methods for patient-specific cardiovascular fluid mechanics modeling},
  author={Takizawa, Kenji and Bazilevs, Yuri and Tezduyar, Tayfun E and Long, Christopher C and Marsden, Alison L and Schjodt, Kathleen},
  journal={Mathematical Models and Methods in Applied Sciences},
  volume={24},
  number={12},
  pages={2437--2486},
  year={2014},
  publisher={World Scientific}
}

@article{kung2014vitro,
  title={In vitro validation of patient-specific hemodynamic simulations in coronary aneurysms caused by Kawasaki disease},
  author={Kung, Ethan and Kahn, Andrew M and Burns, Jane C and Marsden, Alison},
  journal={Cardiovascular Engineering and Technology},
  volume={5},
  number={2},
  pages={189--201},
  year={2014},
  publisher={Springer}
}

@article{sengupta2012image,
  title={Image-based modeling of hemodynamics in coronary artery aneurysms caused by Kawasaki disease},
  author={Sengupta, Dibyendu and Kahn, Andrew M and Burns, Jane C and Sankaran, Sethuraman and Shadden, Shawn C and Marsden, Alison L},
  journal={Biomechanics and Modeling in Mechanobiology},
  volume={11},
  number={6},
  pages={915--932},
  year={2012},
  publisher={Springer}
}

@article{karabelas2018towards,
	title={Towards a computational framework for modeling the impact of aortic coarctations upon left ventricular load},
	author={Karabelas, Elias and Gsell, Matthias AF and Augustin, Christoph M and Marx, Laura and Neic, Aurel and Prassl, Anton J and Goubergrits, Leonid and Kuehne, Titus and Plank, Gernot},
	journal={Frontiers in Physiology},
	volume={9},
	pages={538},
	year={2018},
	publisher={Frontiers}
}

@article{this2019augmented,
	doi = {10.1002/cnm.3223},
	publisher = {Wiley},
	volume = {36},
	number = {3},
	author = {Alexandre This and Ludovic Boilevin-Kayl and Miguel A. Fern{\'{a}}ndez and Jean-Fr{\'{e}}d{\'{e}}ric Gerbeau},
	title = {Augmented resistive immersed surfaces valve model for the simulation of cardiac hemodynamics with isovolumetric phases},
	journal = {International Journal for Numerical Methods in Biomedical Engineering}
}

@article{di2018jet,
  title={Jet collisions and vortex reversal in the human left ventricle},
  author={Di Labbio, Giuseppe and Kadem, Lyes},
  journal={Journal of Biomechanics},
  volume={78},
  pages={155--160},
  year={2018},
  publisher={Elsevier}
}

@article{kilner2000asymmetric,
  title={Asymmetric redirection of flow through the heart},
  author={Kilner, Philip J and Yang, Guang-Zhong and Wilkes, A John and Mohiaddin, Raad H and Firmin, David N and Yacoub, Magdi H},
  journal={Nature},
  volume={404},
  number={6779},
  pages={759--761},
  year={2000},
  publisher={Nature Publishing Group}
}

@article{schwarz2023beyond,
  title={Beyond CFD: Emerging methodologies for predictive simulation in cardiovascular health and disease},
  author={Schwarz, Erica L. and Pegolotti,  Luca and Pfaller,  Martin R. and  Marsden,  Alison L. )},
  journal={Byophysics Reviews},
  volume={4},
  number={1},
  year={2023},
  publisher={AIP Publishing}
}

@article{maceira2006normalized,
  title={Normalized left ventricular systolic and diastolic function by steady state free precession cardiovascular magnetic resonance},
  author={Maceira, Alicia M and Prasad, Sanjay K and Khan, Mohammed and Pennell, Dudley J},
  journal={Journal of {C}ardiovascular Magnetic Resonance},
  volume={8},
  number={3},
  pages={417--426},
  year={2006},
  publisher={Taylor \& Francis}
}

@article{clay2006normal,
  title={Normal range of human left ventricular volumes and mass using steady state free precession MRI in the radial long axis orientation},
  author={Clay, Sarah and Alfakih, Khaled and Radjenovic, Aleksandra and Jones, Timothy and Ridgway, John P},
  journal={Magnetic Resonance Materials in Physics, Biology and Medicine},
  volume={19},
  number={1},
  pages={41--45},
  year={2006},
  publisher={Springer}
}

@article{sugimoto2017echocardiographic,
  title={Echocardiographic reference ranges for normal left ventricular 2D strain: results from the EACVI NORRE study},
  author={Sugimoto, Tadafumi and Dulgheru, Raluca and Bernard, Anne and Ilardi, Federica and Contu, Laura and Addetia, Karima and Caballero, Luis and Akhaladze, Natela and Athanassopoulos, George D and Barone, Daniele and others},
  journal={European Heart Journal-Cardiovascular Imaging},
  volume={18},
  number={8},
  pages={833--840},
  year={2017},
  publisher={Oxford University Press}
}

@article{hammermeister1974rate,
	title={The rate of change of left ventricular volume in man: I. Validation and peak systolic ejection rate in health and disease},
	author={Hammermeister, KE and Brooks, RC and Warbasse, JR},
	journal={Circulation},
	volume={49},
	number={4},
	pages={729--738},
	year={1974},
	publisher={Am Heart Assoc}
}

@article{johnson2008coronary,
  title={Coronary artery flow measurement using navigator echo gated phase contrast magnetic resonance velocity mapping at 3.0 T},
  author={Johnson, Kevin and Sharma, Puneet and Oshinski, John},
  journal={Journal of Biomechanics},
  volume={41},
  number={3},
  pages={595--602},
  year={2008},
  publisher={Elsevier}
}

@article{schiemann2006mr,
  title={MR-based coronary artery blood velocity measurements in patients without coronary artery disease},
  author={Schiemann, M and Bakhtiary, F and Hietschold, V and Koch, A and Esmaeili, A and Ackermann, H and Moritz, A and Vogl, TJ and Abolmaali, ND},
  journal={European Radiology},
  volume={16},
  number={5},
  pages={1124--1130},
  year={2006},
  publisher={Springer}
}

@article{vankan1997finite,
  title={Finite-element simulation of blood perfusion in muscle tissue during compression and sustained contraction},
  author={Vankan, Wilhelmus J and Huyghe, Jacques M and Slaaf, Dick W and Van Donkelaar, Corrinus C and Drost, Maarten R and Janssen, Jan D and Huson, Anthony},
  journal={American Journal of Physiology-Heart and Circulatory Physiology},
  volume={273},
  number={3},
  pages={H1587--H1594},
  year={1997},
  publisher={American Physiological Society Bethesda, MD}
}

@book{hall2020guyton,
  title={Guyton and Hall Textbook of Medical Physiology},
  author={Hall, John E and Hall, Michael E},
  year={2020},
  publisher={Elsevier Health Sciences}
}

@article{maurer2006aortic,
  title={Aortic regurgitation},
  author={Maurer, Gerald},
  journal={Heart},
  volume={92},
  number={7},
  pages={994--1000},
  year={2006},
  publisher={BMJ Publishing Group Ltd}
}

@article{kajander2011clinical,
  title={Clinical value of absolute quantification of myocardial perfusion with 15O-water in coronary artery disease},
  author={Kajander, Sami A and Joutsiniemi, Esa and Saraste, Markku and Pietil{\"a}, Mikko and Ukkonen, Heikki and Saraste, Antti and Sipil{\"a}, Hannu T and Ter{\"a}s, Mika and M{\"a}ki, Maija and Airaksinen, Juhani and others},
  journal={Circulation: Cardiovascular Imaging},
  volume={4},
  number={6},
  pages={678--684},
  year={2011},
  publisher={Am Heart Assoc}
}

@book{quarteroni2019mathematical,
  title={Mathematical modelling of the human cardiovascular system: data, numerical approximation, clinical applications},
  author={Quarteroni, Alfio and Dede', Luca and Manzoni, Andrea and Vergara, Christian and others},
  volume={33},
  year={2019},
  publisher={Cambridge University Press}
}

@article{sun2014computational,
  title={Computational fluid dynamics in coronary artery disease},
  author={Sun, Zhonghua and Xu, Lei},
  journal={Computerized Medical Imaging and Graphics},
  volume={38},
  number={8},
  pages={651--663},
  year={2014},
  publisher={Elsevier}
}

@article{kim2010patient,
  title={Patient-specific modeling of blood flow and pressure in human coronary arteries},
  author={Kim, Hyun Jin and Vignon-Clementel, IE and Coogan, JS and Figueroa, CA and Jansen, KE and Taylor, CA},
  journal={Annals of Biomedical Engineering},
  volume={38},
  pages={3195--3209},
  year={2010},
  publisher={Springer}
}

@article{zhong2018application,
  title={Application of patient-specific computational fluid dynamics in coronary and intra-cardiac flow simulations: Challenges and opportunities},
  author={Zhong, Liang and Zhang, Jun-Mei and Su, Boyang and Tan, Ru San and Allen, John C and Kassab, Ghassan S},
  journal={Frontiers in Physiology},
  volume={9},
  pages={742},
  year={2018},
  publisher={Frontiers Media SA}
}

@article{athani2021two,
  title={Two-phase non-Newtonian pulsatile blood flow simulations in a rigid and flexible patient-specific left coronary artery (LCA) exhibiting multi-stenosis},
  author={Athani, Abdulgaphur and Ghazali, Nik Nazri Nik and Badruddin, Irfan Anjum and Usmani, Abdullah Y and Kamangar, Sarfaraz and Anqi, Ali E and Ahammad, Nandalur Ameer},
  journal={Applied Sciences},
  volume={11},
  number={23},
  pages={11361},
  year={2021},
  publisher={MDPI}
}

@article{di2022prediction,
  title={Prediction of myocardial blood flow under stress conditions by means of a computational model},
  author={Di Gregorio, Simone and Vergara, Christian and Pelagi, Giovanni Montino and Baggiano, Andrea and Zunino, Paolo and Guglielmo, Marco and Fusini, Laura and Muscogiuri, Giuseppe and Rossi, Alexia and Rabbat, Mark G and others},
  journal={European Journal of Nuclear Medicine and Molecular Imaging},
  pages={1--12},
  year={2022},
  publisher={Springer}
}

@article{padula1965obstruction,
  title={Obstruction of the coronary ostia during systole by the aortic valve leaflets},
  author={Padula, Richard T and Camishion, Rudolph C and Bollinger II, Walter F},
  journal={The Journal of Thoracic and Cardiovascular Surgery},
  volume={50},
  number={5},
  pages={683--690},
  year={1965},
  publisher={Elsevier}
}

@article{rabkin2013differences,
  title={Differences in Coronary Blood Flow in Aortic Regurgitation and Systemic Arterial Hypertension Have Implications for Diastolic Blood Pressure Targets: A Systematic Review and Meta-Analysis},
  author={Rabkin, Simon W},
  journal={Clinical Cardiology},
  volume={36},
  number={12},
  pages={728--736},
  year={2013},
  publisher={Wiley Online Library}
}

@article{kume2008mechanism,
  title={Mechanism of increasing systolic coronary flow velocity in patients with aortic regurgitation.},
  author={Kume, Teruyoshi and Kawamoto, Takahiro and Okura, Hiroyuki and Watanabe, Nozomi and Toyota, Eiji and Neishi, Yoji and Sukmawan, Renan and Yamada, Ryotaro and Akasaka, Takashi and Yoshida, Kiyoshi},
  journal={The Journal of Heart Valve Disease},
  volume={17},
  number={1},
  pages={89--93},
  year={2008}
}

@article{taylor2013computational,
  title={Computational fluid dynamics applied to cardiac computed tomography for noninvasive quantification of fractional flow reserve: scientific basis},
  author={Taylor, Charles A and Fonte, Timothy A and Min, James K},
  journal={Journal of the American College of Cardiology},
  volume={61},
  number={22},
  pages={2233--2241},
  year={2013},
  publisher={American College of Cardiology Foundation Washington, DC}
}

@article{papamanolis2021myocardial,
  title={Myocardial perfusion simulation for coronary artery disease: a coupled patient-specific multiscale model},
  author={Papamanolis, Lazaros and Kim, Hyun Jin and Jaquet, Clara and Sinclair, Matthew and Schaap, Michiel and Danad, Ibrahim and van Diemen, Pepijn and Knaapen, Paul and Najman, Laurent and Talbot, Hugues and others},
  journal={Annals of Biomedical Engineering},
  volume={49},
  pages={1432--1447},
  year={2021},
  publisher={Springer}
}

@article{formaggia2003one,
  title={One-dimensional models for blood flow in arteries},
  author={Formaggia, Luca and Lamponi, Daniele and Quarteroni, Alfio},
  journal={Journal of Engineering Mathematics},
  volume={47},
  pages={251--276},
  year={2003},
  publisher={Springer}
}

@book{vlachopoulos2011mcdonald,
  title={McDonald's blood flow in arteries: theoretical, experimental and clinical principles},
  author={Vlachopoulos, Charalambos and O'Rourke, Michael and Nichols, Wilmer W},
  year={2011},
  publisher={CRC press}
}

@book{katz2010physiology,
  title={Physiology of the Heart},
  author={Katz, Arnold M},
  year={2010},
  publisher={Lippincott Williams \& Wilkins}
}

@article{pezzuto2016space,
  title={Space-discretization error analysis and stabilization schemes for conduction velocity in cardiac electrophysiology},
  author={Pezzuto, Simone and Hake, Johan and Sundnes, Joakim},
  journal={International journal for numerical methods in biomedical engineering},
  volume={32},
  number={10},
  pages={e02762},
  year={2016},
  publisher={Wiley Online Library}
}

@article{woodworth2021numerical,
  title={A numerical study on the effects of spatial and temporal discretization in cardiac electrophysiology},
  author={Woodworth, Lucas A and Cans{\i}z, Bar{\i}{\c{s}} and Kaliske, Michael},
  journal={International journal for numerical methods in biomedical engineering},
  volume={37},
  number={5},
  pages={e3443},
  year={2021},
  publisher={Wiley Online Library}
}

@article{vankana1998mechanical,
  title={Mechanical blood-tissue interaction in contracting muscles: a model study},
  author={Vankana, WJ and Huyghe, Jacques M and van Donkelaar, Corrinus C and Drost, Maarten R and Janssen, JD and Huson, A},
  journal={Journal of Biomechanics},
  volume={31},
  number={5},
  pages={401--409},
  year={1998},
  publisher={Elsevier}
}

@article{barnafi2022multiscale,
  title={A Multiscale Poromechanics Model Integrating Myocardial Perfusion and the Epicardial Coronary Vessels},
  author={Barnafi Wittwer, Nicolás Alejandro and Gregorio, Simone Di and Dede', Luca and Zunino, Paolo and Vergara, Christian and Quarteroni, Alfio},
  journal={SIAM Journal on Applied Mathematics},
  volume={82},
  number={4},
  pages={1167--1193},
  year={2022},
  publisher={SIAM}
}

@article{ambrosi2012active,
  title={Active stress vs. active strain in mechanobiology: constitutive issues},
  author={Ambrosi, D and Pezzuto, S28990101312},
  journal={Journal of Elasticity},
  volume={107},
  pages={199--212},
  year={2012},
  publisher={Springer}
}

@article{smith2002anatomically,
  title={An anatomically based model of transient coronary blood flow in the heart},
  author={Smith, NP and Pullan, AJ and Hunter, Peter J},
  journal={SIAM Journal on Applied mathematics},
  volume={62},
  number={3},
  pages={990--1018},
  year={2002},
  publisher={SIAM}
}

@article{augustin2016anatomically,
  title={Anatomically accurate high resolution modeling of human whole heart electromechanics: a strongly scalable algebraic multigrid solver method for nonlinear deformation},
  author={Augustin, Christoph M and Neic, Aurel and Liebmann, Manfred and Prassl, Anton J and Niederer, Steven A and Haase, Gundolf and Plank, Gernot},
  journal={Journal of computational physics},
  volume={305},
  pages={622--646},
  year={2016},
  publisher={Elsevier}
}

@article{augustin2016patient,
  title={Patient-specific modeling of left ventricular electromechanics as a driver for haemodynamic analysis},
  author={Augustin, Christoph M and Crozier, Andrew and Neic, Aurel and Prassl, Anton J and Karabelas, Elias and Ferreira da Silva, Tiago and Fernandes, Joao F and Campos, Fernando and Kuehne, Titus and Plank, Gernot},
  journal={EP Europace},
  volume={18},
  number={suppl\_4},
  pages={iv121--iv129},
  year={2016},
  publisher={Oxford University Press}
}


\section*{Acknowledgements}
A.Z., C.V., L.D., and A.Q. received funding from the Italian Ministry of University and Research (MIUR) within the PRIN (Research projects of relevant national interest 2017 “Modeling the heart across the scales: from cardiac cells to the whole organ” Grant Registration number 2017AXL54F). 

A.Z., C.V., L.D., F.R., and A.Q. are members of the INdAM group GNCS ``Gruppo Nazionale per il Calcolo Scientifico'' (National Group for Scientific Computing). This work has been supported by the Italian GNCS under the INdAM GNCS Project 
\\
CUP\textunderscore E55F22000270001.

The authors acknowledge the CINECA award under the ISCRA initiative, for the availability of high performance computing resources and support under the projects IsC87\_MCH, P.I. A. Zingaro, 2021-2022 and  IsB25\_MathBeat, P.I. A. Quarteroni, 2021-2022.

\section*{Author contributions statement}
A.Z.: Conceptualization,  methodology, software implementation, simulation, formal analysis, writing (original draft). C.V.: Project administration, conceptualization, methodology, formal analysis, writing (review). L.D.: Methodology, formal analysis. F.R.: Methodology, formal analysis. A.Q.: Funding acquisition, project administration, conceptualization, methodology. All authors edited the manuscript.

\section*{Additional information}

\textbf{Accession codes} The datasets used and/or analysed during the current study are available from the corresponding author on reasonable request; 
\\
\textbf{Competing interests}: The authors declare no competing interests. 


\end{document}